\documentclass{article}

\PassOptionsToPackage{numbers, compress}{natbib}
\usepackage{color,soul}

 \usepackage[preprint]{neurips_2026}

\usepackage[utf8]{inputenc} 
\usepackage[T1]{fontenc}    
\usepackage{hyperref}       
\usepackage{url}            
\usepackage{booktabs}       
\usepackage{amsfonts}       
\usepackage{nicefrac}       
\usepackage{microtype}      
\usepackage{xcolor}         

\usepackage{multirow}
\usepackage{hyperref}
\usepackage{microtype}
\usepackage{graphicx}
\usepackage{subcaption}
\usepackage{booktabs} 
\usepackage{amsmath}
\usepackage{amssymb}
\usepackage{wrapfig}
\usepackage{placeins}
\usepackage{algorithm}
\usepackage{algpseudocode}
\usepackage{todonotes}

\usepackage{arydshln}
\usepackage{colortbl}

\title{Predicting directional flexibility in proteins}

\author{%
Vsevolod Viliuga
\thanks{Corresponding author. Contact: vsevolod.viliuga@scilifelab.se}\,\,\,$^{1,2}$
\quad
Leif Seute\,\,\,$^{2,3,4}$
\quad
Matteo Tadiello\,$^{1}$
\quad
Nicolas Wolf\,\,\,$^{2,3,4}$ \\
\textbf{Frauke Gräter}\,$^{2,4}$
\quad
\textbf{Arne Elofsson}\,$^{1}$ \\
\\$^1$ DBB and SciLifeLab at Stockholm University, Stockholm, Sweden\\
$^2$Max Planck Institute for Polymer Research, Mainz, Germany\\
$^3$Heidelberg Institute for Theoretical Studies, Heidelberg, Germany\\
$^4$IWR, Heidelberg University, Heidelberg, Germany\\
}

\begin{document}

\maketitle

\begin{abstract}

Predicting protein dynamics is a long-standing problem in computational structural biology. Often, protein function critically depends on local directed motions, such as hinge movements, catalytic loop rearrangements and domain reorientations, which can be characterized by directional flexibility and correlated structural motions of the protein backbone. While Molecular Dynamics (MD) simulations provide an established but often prohibitively expensive approach, recent deep generative models aim to reduce this cost by directly predicting conformational ensembles, emulating MD. However, due to their large size and the need to generate several states until the derived dynamical properties converge, these models remain expensive. In this work, we propose BackFlip-2: a fast SE(3)-equivariant graph neural network trained to directly predict dynamical descriptors, such as directional backbone flexibility and pairwise dynamic correlations, from an equilibrium structure. In a series of experiments, we show that our model matches the accuracy of substantially larger ensemble generation models while being orders of magnitude faster, and demonstrate that the proposed equivariant architecture is especially well-suited for capturing anisotropic motions in proteins. BackFlip-2 model weights, training and inference code are available at \url{https://github.com/graeter-group/backflip}.
\end{abstract}
\section{Introduction}
\label{sec:introduction}

Proteins are essential biomolecules that carry out a wide range of functions, ranging from DNA replication and energy metabolism to sophisticated coordinated processes such as chromosome segregation and cell division~\cite{voetvoet}. Since the three-dimensional structure of a protein determines much of its function, accurately describing structural features is central to understanding its biological role. Importantly, proteins are not static objects: their function can depend on both large conformational changes occurring on timescales beyond milliseconds, but also on local dynamical degrees of freedom explored within hundreds of nanoseconds. Such local, fast motions can coordinate protein assembly and oligomerization~\cite{khmelinskaia2025local}, shape active site geometry~\cite{schwartz2023protein, nam2023protein, du2025conformational} and determine substrate specificity~\cite{huang2025novo}. In turn, larger and typically slower loop, hinge, and domain motions drive catalysis~\cite{tousignant2004protein, kurkcuoglu2012coupling, schwartz2023protein}, molecular recognition~\cite{spyrakis2011protein, csermely2010induced}, and allosteric regulation~\cite{reiss2020allosteric}. 
Importantly, many of these functional motions are anisotropic~\cite{teilum2009functional, zhou2014aligning}, i.e. the protein structural elements involved in the function move along a preferred direction. For example, in ionotropic glutamate receptors, ligand-binding domain closure is coupled to channel gating~\cite{twomey2018structural}, while coordinated directional LID and NMPbind domain movements in adenylate kinase guide catalysis~\cite{galenkamp2024allostery, ping2013molecular}. Thus, understanding protein function requires a description of protein dynamics across different timescales, capturing directionality and collective motions.

An established way to characterize protein dynamics is to run Molecular Dynamics (MD) simulations. In MD, a molecular system is evolved in time by numerically integrating Newton’s equations of motion. However, numerical stability requires femtosecond-scale integration time steps, while biologically relevant processes often emerge only over much longer timescales, rendering MD computationally expensive and hence limiting its scalability to large sets of proteins. A faster physics-based alternative is Normal Mode Analysis (NMA), which describes protein motions around an equilibrium structure assuming an effective harmonic potential \cite{brooks1983harmonic}.
However, this approximation is restricted to strictly local fluctuations \cite{hinsen2005normal, hayward2008normal} and often suffers from limited accuracy \cite{diepeveen_2024}.

\paragraph{Ensemble generation models}
Recently, several deep generative models based on diffusion~\cite{ho2020ddpm} and flow matching~\cite{lipman2022flow,albergo2022building} have been proposed to obtain protein conformational ensembles at reduced cost by emulating MD simulation. Reflecting the importance of dynamics across timescales, different models have targeted different time horizons. For example, BioEmu~\cite{bioemu} focuses on long, millisecond-scale conformational ensembles, while AlphaFlow~\cite{alphaflow}, BBFlow~\cite{bbflow}, and related methods~\cite{wang2024proteinconfdiff,jing2024mdgen} model MD ensembles on shorter, sub-microsecond timescales by training on the ATLAS dataset of 300 ns MD trajectories~\cite{vander2024atlas}. These approaches accurately emulate MD-like conformational ensembles, but this comes at a cost: the ensemble generation models are typically large and require the sampling of multiple conformations to obtain converged dynamical observables, with inference times ranging from several minutes to hours for a single protein.


\paragraph{Predicting dynamical descriptors directly}
In a complementary line of work, structural flexibility is predicted directly, avoiding ensemble generation altogether. Models such as Flexpert~\cite{flexpert}, Pegasus~\cite{vander2025pegasus}, and BackFlip~\cite{flips} predict MD-derived per-residue flexibility from sequence or structure. However, these models predict scalar isotropic flexibility (e.g. RMSF). Yet, most of the functional motions are characterized not only by the extent of residue movements, but also by their direction and by residue-residue correlations \cite{teilum2009functional}. The recently proposed DynaProt~\cite{dynaprot} attempts to address this limitation by predicting per-residue covariance matrices describing anisotropic flexibility, together with pairwise residue couplings. However, the covariance matrices predicted by DynaProt are SE(3)-invariant, which is an important drawback: covariance matrices derived from MD trajectories are not invariant but transform according to the matrix representation of \(R \in \mathrm{SO}(3)\) under global rotation.

\paragraph{Main contributions}
In this work, we propose BackFlip-2: an SE(3)-equivariant neural network for predicting per-residue directional flexibility on the 300-500 ns timescale regime from the protein backbone of an equilibrium structure. BackFlip-2 predicts anisotropic covariance matrices and pairwise residue correlations, generalizing our previous direction agnostic model BackFlip~\cite{flips}.
The proposed model achieves state-of-the-art performance in both accuracy and runtime. It outperforms directly comparable flexibility prediction baselines and is even on par with substantially larger ensemble generative models, such as AlphaFlow~\cite{alphaflow}, while being over four orders of magnitude faster.
We achieve this by constructing a lightweight model based on the IPA algorithm from AlphaFold2 \cite{alphafold2} for predicting covariances that transform under the matrix representation of $\mathrm{SE}(3)$, and scalar pairwise residue couplings. We train and evaluate our model on the well-established ATLAS dataset~\cite{vander2024atlas} comprising MD simulations for 1390 proteins each covering in total 300 ns, and further evaluate it on mdCATH \cite{mdcath} and \textit{de novo} proteins. Our ablations demonstrate that due to the proposed equivariance, the model performs especially well at capturing highly anisotropic residue motions with directional specificity, which are of particular interest in practice.

\subsection{Related Work}
\label{sec:related_work}

\paragraph{MD simulation datasets}
The ATLAS dataset~\cite{vander2024atlas} provides standardized MD simulations for 1390 structurally non-redundant proteins, each with three independent 100 ns replicas obtained with the CHARMM36m force-field at 300 K and a 2 fs integration time step \cite{vander2024atlas}. The mdCATH dataset \cite{mdcath} provides MD simulations for 5398 protein domains taken from the CATH database \cite{orengo1999cath} with five trajectories per domain simulated at different temperatures (320, 348, 379, 413, and 450 K) covering 500 ns in total, using the CHARMM22 force field and a 4 fs integration time step.

\paragraph{Predicting protein dynamics}
An early approach to predict protein dynamics faster than MD is Normal Mode Analysis (NMA), which approximates local perturbations of the protein by an effective harmonic potential, limiting it to small-amplitude motions~\cite{hinsen2005normal, hayward2008normal}. 
Flow- and diffusion-based generative models instead learn conformational ensembles from MD data: In AlphaFlow~\cite{alphaflow}, the protein structure prediction model AlphaFold2~\cite{alphafold2} is fine-tuned as a flow matching model on MD trajectories from ATLAS, significantly outperforming NMA and also the training-free MSA subsampling approach proposed by~\citet{wayment-steele2024predicting}. BBFlow~\cite{bbflow} formulates ensemble generation as a structure-to-structure flow matching task and is substantially faster than AlphaFlow. 

To directly predict RMSF, several model have been proposed previously (see also App.~\ref{sec:app:flexibility_baselines}):
FlexPert~\cite{flexpert} and Pegasus~\cite{vander2025pegasus} predict per-residue RMSF from the sequence using pretrained protein language models, while the faster BackFlip~\cite{flips} predicts isotropic RMSF from the protein structure using a graph neural network. The recent model DynaProt~\cite{dynaprot} generalizes the idea from BackFlip to predict flexibility based on the protein backbone structure beyond RMSF and proposes to predict per-residue $3\times 3$ covariance matrices and pairwise residue couplings with an invariant graph neural network.

\section{Per-residue covariances and pairwise couplings as directional flexibility}
\label{sec:targets_description}

We aim to learn two observables from MD simulations characterizing residue-wise dynamical degrees of freedom and residue correlations:
\emph{per-residue covariances}
$\Sigma^{(ii)}\in\mathbb{R}^{3\times3}$,
which describe anisotropic fluctuations of residue $i$, and \emph{pairwise scalar couplings} $C\in\mathbb{R}^{N\times N}$, which capture correlated motions between residue pairs. We will define these quantities below.
\paragraph{Full covariance matrix}
To compute $\Sigma^{(ii)}$ and $C$ from MD trajectories, we first superpose all trajectory frames onto an \emph{equilibrium reference conformation} to remove global translation and rotation.
The full positional covariance $\Sigma_\text{full}$ is then defined as covariance of the stacked positions $x\in\mathbb{R}^{3N}$,
\begin{equation}
    \Sigma_{\mathrm{full}}
    \equiv
    \big\langle
    \left(x - \langle x \rangle\right)
    \otimes
    \left(x - \langle x \rangle\right)
    \big\rangle
    \in \mathbb{R}^{3N \times 3N},
    \label{eq:full_cov}
\end{equation}
where $\langle \rangle$ denotes the ensemble average and $\otimes$ is the tensor product.

The full covariance matrix can be decomposed into $N^2$ individual $3\times 3$ matrices describing the correlation between residues $i$ and $j$,
\begin{equation}
    \Sigma^{(ij)} = 
    \big\langle
    \left(x_i - \langle x_i \rangle\right)
    \otimes
    \left(x_j - \langle x_j \rangle\right)
    \big\rangle
    \in \mathbb{R}^{3 \times 3}.
    \label{eq:per-res-covariance-definition}
\end{equation}
Crucially, under global rotations $R\in\mathrm{SO}(3)$ of the reference structure, these matrices transform under the matrix representation of $\mathrm{SO}(3)$:
\begin{equation}
    \Sigma^{(ij)} \mapsto
    \big\langle
    \left(R\,x_i - \langle R\,x_i \rangle\right)
    \otimes
    \left(R\,x_j - \langle R\,x_j \rangle\right)
    \big\rangle
    =
    R\,\Sigma^{(ij)}R^T,
    \label{eq:cov_trafo_behaviour}
\end{equation}

\paragraph{Per-residue covariance}
We then define the \emph{per-residue covariance} $\Sigma$ as subset of the covariance matrices in Eq. \ref{eq:per-res-covariance-definition},
$
    \Sigma
    \equiv
    \{\Sigma^{(ii)}\}_{i=1}^N,
$
which can be interpreted as measure for directional flexibility (anisotropy) of a residue (visualization in Fig.~\ref{fig:exp_ellipsoids}).
From these per-residue covariances, one can obtain the scalar \emph{root-mean-square fluctuations} (RMSF) by taking the trace,
\begin{equation}
\mathrm{RMSF}_i
=
\sqrt{
\left\langle
\sum_{d=1}^{3}
\left(
x_{i,d}-\left\langle x_{i,d}\right\rangle
\right)^2
\right\rangle
}
=
\sqrt{
\sum_{d=1}^{3}
\left\langle
\left(
x_{i,d}-\left\langle x_{i,d}\right\rangle
\right)^2
\right\rangle
}
=
\sqrt{
\sum_{d=1}^{3}
\Sigma^{(ii)}_{dd}
}
=
\sqrt{
\mathrm{tr}\!\left(\Sigma^{(ii)}\right)
},
\label{eq:rmsf_from_cov}
\end{equation}
where we used the linearity of the ensemble average operator $\langle\,\rangle$ in the second equality.
In this sense one can think of the per-residue covariance as a generalization of the established scalar RMSF to quantify the anisotropy of fluctuations.
Note that the RMSF does not change under global rotations since the trace is $\mathrm{O}(n)$ invariant,
\begin{equation}
\forall R \in \mathrm{O}(n),\ \forall A \in \mathbb{R}^{n\times n}:
\quad
\mathrm{tr}\!\left(R A R^T\right)
=
\mathrm{tr}(A).
\label{eq:trace_is_invariant}
\end{equation}

\paragraph{Pairwise residue couplings}
We define the \emph{pairwise residue couplings} between residues $i$ and $j$ as the trace of the off-diagonal block matrices $\Sigma^{(ij)}$,
\begin{equation}
\label{eq:couplings}
    C_{ij} \equiv \mathrm{tr}\left(\Sigma^{(ij)}\right)
    =
    \big\langle
    \left(x_i-\langle x_i\rangle\right)
    \cdot
    \left(x_j-\langle x_j\rangle\right)
    \big\rangle
    ,
\end{equation}
that is as the correlation of the relative displacements of both residues across the ensemble.
These couplings hence indicate how likely it is that two residues move in the same direction (positive values of $C_{ij}$) or antiparallel (negative values of $C_{ij}$).
Importantly, the trace is rotation invariant (Eq.~\ref{eq:trace_is_invariant}), which makes the couplings independent of global rotations of the reference structure.
This is in contrast to the definition in~\cite{dynaprot}, where the mean is used instead of the trace, making the couplings dependent on the absolute orientation of the reference structure. From \(\mathrm{C}\), one can compute the established \emph{dynamic cross-correlation matrix} (DCCM) \(\tilde{\mathrm{C}}\) by normalizing each entry with the corresponding per-residue fluctuation magnitudes:
\begin{equation}
\label{eq:dccm}
\tilde{\mathrm{C}} \in \mathbb{R}^{N \times N},
\qquad
\tilde{\mathrm{C}}_{ij}
=
\frac{\mathrm{C}_{ij}}{\sqrt{\mathrm{C}_{ii}\,\mathrm{C}_{jj}}}
=
\frac{\mathrm{C}_{ij}}{\mathrm{RMSF}_i\,\mathrm{RMSF}_j}.
\end{equation}
The resulting entries satisfy \(\tilde{\mathrm{C}}_{ij} \in [-1,1]\) and quantify the correlation of displacement directions.

\section{Method}


We propose an $\mathrm{SE}(3)$-equivariant network that predicts (i) equivariant per-residue covariances and (ii) scalar pairwise residue couplings, as derived from MD simulations. The model relies on the protein backbone geometry and does not require evolutionary sequence information or pre-trained protein language model embeddings, enabling orders of magnitude faster inference at improved accuracy. It consists of two components: a backbone encoder and a dynamics prediction module that
receives learned backbone representations. For a structure with $N$ residues, the model jointly predicts
\begin{equation}
    (\hat{\Sigma}, \hat{C}) \in \mathbb{R}^{N \times 3 \times 3} \times \mathbb{R}^{N \times N},
\end{equation}
with the per-residue covariance $\Sigma$ and pairwise coupling $C$ as defined above.
For training, we define a composite objective $\mathcal{L} = \mathcal{L}_{\Sigma} + \mathcal{L}_{C}$ (see Eq. \ref{eq:L_sigma} and \ref{eq:L_C}). In the following, we describe the model architecture in detail.

\subsection{Backbone encoder}
We follow the well-established representation of a protein backbone as a set of rigid body poses 
$T \in \mathrm{SE}(3)^{N}$,
where each residue frame is given by 
$T_i = (x_i, R_i)$
with translational component $x_i \in \mathbb{R}^3$ and rotational component 
$R_i \in \mathrm{SO}(3)$.
To learn an expressive representation of protein backbone geometry, we make use of the Invariant Point Attention (IPA) introduced in AlphaFold2~\cite{alphafold2}.
The intermediate residue features $h \in \mathbb{R}^{N \times d}$ express each node embedding $h_i$ in the local coordinate system defined by the residue frame $T_i$.
Upon message passing, the node features are transformed into the coordinate frame of the target residue, as in AlphaFold2.
This ensures $\mathrm{SE}(3)$ equivariance under arbitrary node-wise learnable layers, and also allows the model to capture geometric information about relative displacements and orientations between residues. We initialize $h_i$ by combining an amino acid type one-hot-encoding with a positional encoding of the index in the protein chain, followed by a projection to the latent dimension $d$.
Initial edge features $e \in \mathbb{R}^{N \times N \times c}$ are defined for all residue pairs by combining a relative positional index encoding with the pairwise distance between the corresponding residues. Pairwise $C_{\alpha}$ distances are binned uniformly between 0 and 20\, \AA. Node and edge embeddings are refined by a stack of $l$ modified IPA blocks. Each block constructs $\mathrm{SE}(3)$-invariant attention scores from point-valued query and key features, followed by a transformer acting on the scalar node features~\cite{alphafold2}.
Unlike structure prediction models, the backbone frames are held fixed, i.e. the backbone update block is omitted, only node and edge embeddings are updated in each layer.
After $l$ blocks, both the node embeddings $h^{(l)}$ and edge embeddings $e^{(l)}$ are passed to the prediction heads described below.
We set $d = 96$, $c = 64$, and $l = 4$.

\subsection{Dynamics prediction module}
\label{sec:dynamics_prediction_module}

\begin{figure}
    \centering
    \includegraphics[width=.9\textwidth]{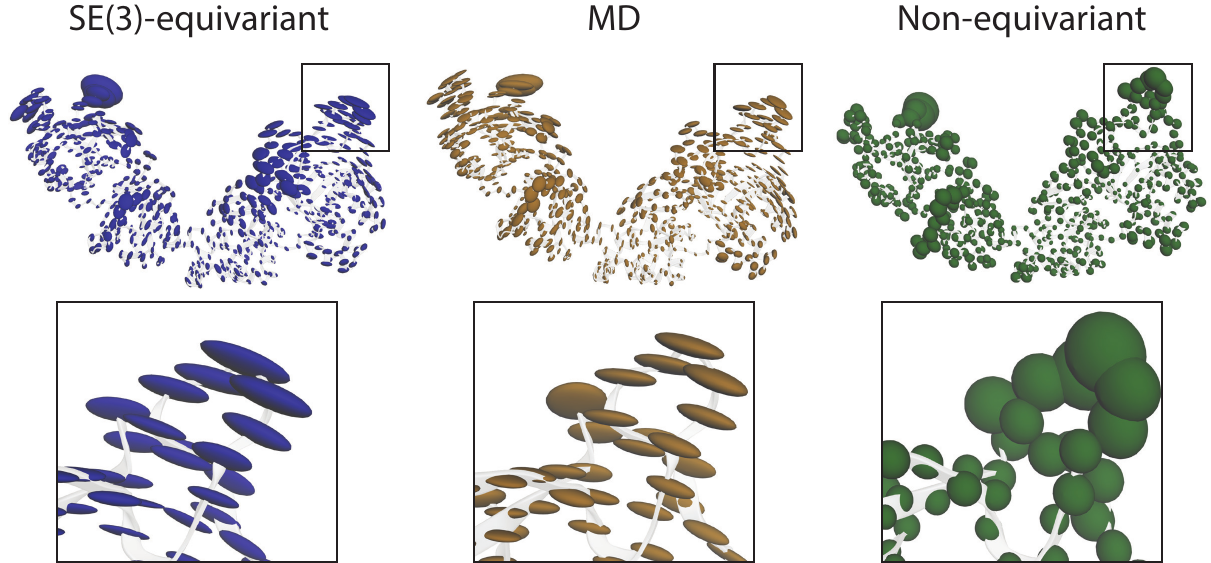}
    \vspace{0.1cm}
    \caption{Equivariance is crucial to capture anisotropic covariances. We visualize ellipsoids representing per-residue covariance matrices for the ATLAS test set protein 7asgA predicted with our proposed $\mathrm{SE}(3)$-equivariant model (left), the MD-derived ground truth (middle) and the prediction of the non-equivariant model variant from Sec.~\ref{sec:anisotropy} (right).}
    \label{fig:exp_ellipsoids}
\end{figure}

We propose a dynamics prediction module that maps the learned backbone node and edge embeddings to per-residue covariances and pairwise residue couplings, respectively. A key architectural difference to the recently proposed DynaProt~\cite{dynaprot} is that we predict $\mathrm{SE}(3)$-equivariant per-residue covariance matrices according to the transformation behavior derived in Eq.~\ref{eq:cov_trafo_behaviour}, whereas DynaProt proposes an $\mathrm{SE}(3)$-invariant covariance head. In our experiments we demonstrate that the $\mathrm{SE}(3)$ equivariance is crucial for capturing anisotropies in the covariances (Fig.~\ref{fig:exp_ellipsoids} and Sec. \ref{sec:anisotropy}). Beyond this, our dynamics prediction module leverages the edge embeddings $e^{(l)}$ returned by the backbone encoder for the prediction of pairwise residue couplings, which encode pairwise geometric information and interaction strengths between residues directly. DynaProt instead predicts both quantities from node embeddings alone, and introduces a separate pair-attention module to capture inter-residue dependencies. The use of edge encodings eliminates the need for a heavy pair-attention module, substantially reducing parameter count while also improving predictive performance, as we demonstrate in our experiments. We describe both heads in detail below.

\subsubsection{Equivariant per-residue covariance head}

To predict the per-residue covariance matrices $\Sigma^{(ii)}$, we construct an output head that satisfies two constraints by design: symmetric positive semi-definiteness, and $\mathrm{SE}(3)$ equivariance.
We use this as inductive bias since the ground truth covariances satisfy both of these constraints (Eq.~\ref{eq:per-res-covariance-definition} and Eq.~\ref{eq:cov_trafo_behaviour}).

\paragraph{Symmetric positive semi-definiteness}
We employ an Multi-Layer Perceptron (MLP) that maps the node embedding $h_i^{(l)}$ to a matrix $A^{(i)} \in \mathbb{R}^{3 \times 3}$, from which we construct the predicted local-frame covariance matrix as
\begin{equation}
    \hat{\Sigma}^{(ii)}_\mathrm{loc} = A^{(i)}{A^{(i)}}^\top,
    \label{eq:local_cov}
\end{equation}
which by construction is symmetric positive semi-definite (SPSD) for any matrix $A^{(i)}$.
We compare our approach to the Cholesky-factorization method from DynaProt in App.~\ref{sec:app_cholesky_comparison}.

\paragraph{SE(3) equivariance}
Since the target blocks $\Sigma^{(ii)}$ computed from MD trajectories transform equivariantly under global rotations (Eq.~\ref{eq:cov_trafo_behaviour}), the predicted covariance should also satisfy this behavior. In order to obtain an equivariant predicted covariance matrix, we rotate the local covariance into the global frame using the residue rotation $R_i \in \mathrm{SO}(3)$,
\begin{equation}
    \hat{\Sigma}^{(ii)} = R_i\, \hat{\Sigma}^{(ii)}_\mathrm{loc}\, R_i^\top.
    \label{eq:cov_head}
\end{equation}
Under a global rotation $R \in \mathrm{SO}(3)$, the residue frames transform as $R_i \mapsto R R_i$, while $\hat{\Sigma}^{(ii)}_\mathrm{loc}$ is unchanged since it is defined locally.
The predicted global-frame covariance therefore transforms as
\begin{equation}
    \hat{\Sigma}^{(ii)} \;\mapsto\; (RR_i)\,\hat{\Sigma}^{(ii)}_\mathrm{loc}\,(RR_i)^\top = R\,\hat{\Sigma}^{(ii)}\,R^\top,
\end{equation}
matching the required equivariance from Eq.~\ref{eq:cov_trafo_behaviour}.
This rotated matrix is also SPSD by definition since
\begin{equation}
    \forall v \in \mathbb{R}^3:\quad
    v^\top \hat{\Sigma}^{(ii)} v
    =
    v^\top \left(R_i\,\hat{\Sigma}^{(ii)}_\mathrm{loc}\,R_i^\top \right) v
    =
    (R_i^\top v)^\top \hat{\Sigma}^{(ii)}_\mathrm{loc}\, (R_i^\top v) \geq 0,
\end{equation}
where we used SPSD of $\hat{\Sigma}^{(ii)}_\mathrm{loc}$ in the last step.
Hence the constructed learnable covariance head $\hat{\Sigma}^{(ii)}$ satisfies both of the constraints proposed above.
We show in our ablation study in Sec.~\ref{sec:ablation} that both of these constraints are indeed favorable in practice, especially for anisotropic covariances (Fig.~\ref{fig:exp_ellipsoids}).

\paragraph{Loss function}
As in DynaProt~\cite{dynaprot}, we train the covariance head with a log-Frobenius loss between the predicted and ground-truth covariance matrices, given by
\begin{equation}
\mathcal{L}_{\Sigma}
=
\left\|
\log\!\left(\hat{\Sigma}^{(ii)}\right)
-
\log\!\left(\Sigma^{(ii)}\right)
\right\|_F^2,
\label{eq:L_sigma}
\end{equation}
where $\log(\cdot)$ denotes the matrix logarithm and $\|\cdot\|_F$ the Frobenius norm.

\subsubsection{Pairwise residue couplings head}

The head for predicting pairwise residue couplings takes as input the backbone edge representation $e^{(l)} \in \mathbb{R}^{N \times N \times c}$ and predicts a pairwise scalar coupling matrix $\hat{C} \in \mathbb{R}^{N \times N}$.
The target couplings computed from MD trajectories are invariant under global rotations (Eq.~\ref{eq:couplings}), hence we construct the predicted couplings from the edge embeddings $e_{ij}^{(l)}$, which are $\mathrm{SE}(3)$-invariant scalar features.
For each residue pair $(i,j)$, we directly pass the edge embedding $e_{ij}^{(l)}$ through an MLP to predict $\hat{C}_{ij}$.
We symmetrize the predicted coupling matrix as $\hat{C} = (\hat{C} + \hat{C}^\top)/2$ to ensure that the predicted coupling matrix is symmetric.

\paragraph{Loss function}
We train the coupling head with a composite loss, with the primary objective being the correlation matrix distance (CMD) loss~\cite{cmd}:
\begin{equation}
d_{\mathrm{corr}}(\hat{C}, C)
=
1 - \frac{\mathrm{tr}(\hat{C}\, C)}{\|\hat{C}\|_F \, \|C\|_F}.
\end{equation}
CMD measures the similarity between two coupling matrices through their normalized Frobenius inner product.
However, because it is scale-invariant, we complement it with an MAE term on the predicted coupling values to explicitly enforce the correct magnitude. In addition, we analytically compute the normalized DCCM as in Eq.~\ref{eq:dccm} induced by the predicted pairwise scalar coupling matrix and apply an auxiliary MSE loss on it.
The composite loss function is then given as
\begin{equation}
\mathcal{L}_{C}
=
d_{\mathrm{corr}}(\hat{C}, C)
+
\lambda_{\mathrm{MAE}}\, \mathrm{MAE}(\hat{C}, C)
+
\lambda_{\mathrm{DCCM}}\, \mathrm{MSE}\!\left(\hat{\tilde{C}}, \tilde{C}\right),
\label{eq:L_C}
\end{equation}
where $\hat{C}$ denotes the predicted and $C$ the ground-truth coupling matrix, $\hat{\tilde{C}}$ and $\tilde{C}$ the corresponding normalized DCCMs, and $\lambda_{\mathrm{MAE}}, \lambda_{\mathrm{DCCM}}$ are hyperparameters denoting loss weights.
\newcommand{\error}[3]{{#1}^{\, #2}_{\, #3}} 

\section{Experiments}
\label{sec:experiments}

\paragraph{Training}

We train BackFlip-2 on the ATLAS dataset using both the AlphaFlow \cite{alphaflow} and FlexPert \cite{flexpert} data splits and compare against baselines on their respective split for fairness. We train the model and all ablation variants \emph{jointly} on both per-residue covariance and pairwise coupling tasks for 200 epochs, requiring approximately 4 GPU hours on a single NVIDIA A100 GPU, and select the checkpoints based on the lowest validation loss.
We additionally construct and train a model on a combined ATLAS/mdCATH dataset to assess the effect of joint training on MD data obtained under different simulation settings, following the same procedure as for ATLAS (details in App.~\ref{sec:app_ATLAS_mdCATH_dataset}). All datasets along with the training, validation and test splits, and model weights are available at \url{https://github.com/graeter-group/backflip}.

\paragraph{Metrics}

We evaluate how well the model predicts three MD-derived dynamical descriptors: (i) per-residue covariances (Eq.~\ref{eq:per-res-covariance-definition}), (ii) RMSF derived from the predicted covariances via Eq.~\ref{eq:rmsf_from_cov}, and (iii) pairwise residue couplings (Eq.~\ref{eq:couplings}).
For per-residue covariance, we assess accuracy of the predictions by reporting the variance contribution of the root mean Wasserstein distance (RMWD) between predicted and MD covariance matrices, together with the variance contribution to the symmetric KL divergence, both proposed in AlphaFlow~\cite{alphaflow}. To explicitly evaluate the directionality of the predicted covariances, we construct ellipsoids from the predicted and MD per-residue covariances and report their volume overlap using IoU and Dice metrics, as described in App.~\ref{sec:app_metrics}.
For RMSF, we report the per-target Pearson correlation coefficient $r$ and mean absolute error (MAE) to the ground truth. For evaluating the pairwise residue couplings, we reconstruct the induced dynamic cross-correlation matrices (DCCM, Eq.~\ref{eq:dccm}) as in~\cite{bbflow} and report Pearson correlation and MAE over all residue pairs. All ground-truth quantities are computed from three stacked MD trajectories.

\paragraph{Baselines}
For the RMSF prediction task, we compare BackFlip-2 to FlexPert, Pegasus, its predecessor BackFlip, DynaProt, AlphaFlow-MD-Templates (AFMD-T), BBFlow, and BioEmu (Sec.~\ref{sec:related_work}).
FlexPert, Pegasus and BackFlip are trained to predict RMSF directly whereas our model and DynaProt predict per-residue covariance, from which we compute RMSF analytically via Eq.~\ref{eq:rmsf_from_cov}.
AFMD-T, BBFlow, and BioEmu generate conformational ensembles, from which we compute RMSF.
For per-residue covariance and residue couplings prediction, we compare our model to DynaProt, AFMD-T, BBFlow, and BioEmu.
We note that BioEmu was not trained on the standardized ATLAS dataset, but generates ensembles corresponding to MD-timescales way beyond the 300 ns timescale of ATLAS (Sec.~\ref{sec:related_work}), explaining its poor performance if evaluated on ATLAS. We include it for completeness. For DynaProt, neither code or model weights are published at the time of submission, hence we can only compare the values reported in the paper.

\subsection{ATLAS benchmark performance}


\paragraph{Predicting RMSF}

\begin{figure*}
    \centering
    \includegraphics[width=.9\textwidth]{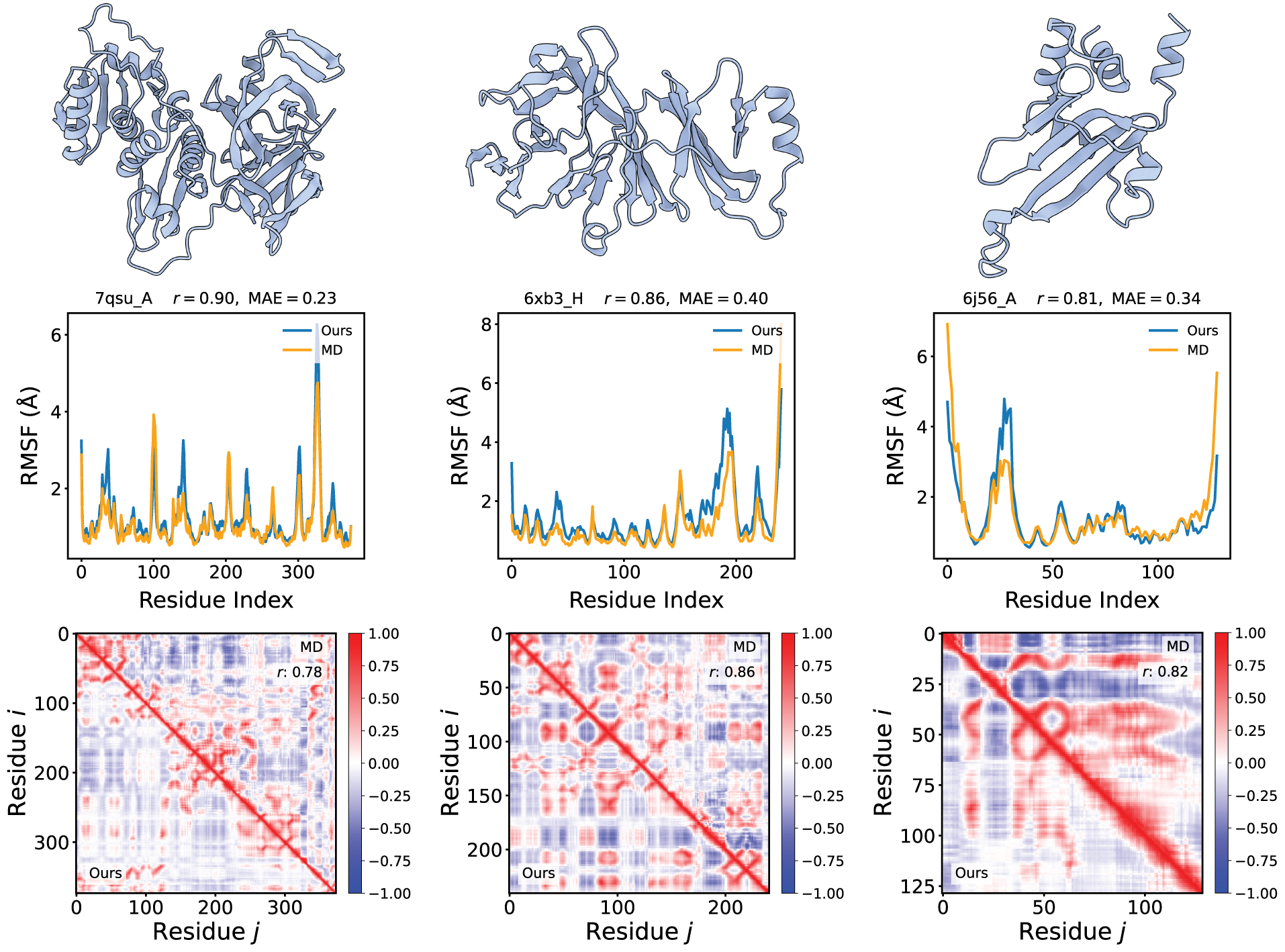}
    \caption{Predictions of per-residue RMSF and pairwise residue couplings (DCCM) made by BackFlip-2 for three proteins from the ATLAS test set compared to MD. For the DCCM matrices, lower triangle of the matrix is the prediction, and upper is MD.}
    \label{fig:exp_rmsf_couplings}
\end{figure*}

The proposed model outperforms RMSF prediction models FlexPert, Pegasus and the predecessor BackFlip on RMSF prediction task. On the held-out ATLAS test set, it achieves higher per-target correlation and lower MAE with respect to MD than all baselines (Tab.~\ref{tab:rmsf_performance}). Compared to FlexPert and Pegasus, BackFlip-2 is also at least an order of magnitude faster and uses substantially fewer trainable parameters. BackFlip is similarly efficient, but remains less accurate. In addition to the ATLAS test set, we evaluate RMSF prediction on 100 \textit{de novo} proteins generated with RFdiffusion~\cite{watson2023novo} and FrameFlow~\cite{yim2023frameflow} and simulated under the same settings as ATLAS (made available in~\cite{bbflow}). Importantly, our proposed model remains robust for \textit{de novo} proteins: it also achieves state-of-the-art performance on this dataset. The performance of FlexPert and Pegasus, which rely on protein Language Model embeddings, however, deteriorates substantially. We visualize RMSF predictions for three randomly selected ATLAS test proteins in Fig.~\ref{fig:exp_rmsf_couplings}.
\begin{table*}
    \centering
    \caption{
    BackFlip-2 achieves state-of-the-art performance among RMSF prediction models.
    We evaluate all models on the ATLAS test set (choosing the FlexPert split \cite{flexpert} for a fair comparison) and on 100 \textit{de novo} proteins.
    We report the per-target Pearson correlation \(r\) and mean absolute error (MAE) between predicted and ground-truth RMSF.
    Metrics are reported as \textit{mean} over proteins in each test set with deviation computed by bootstrapping samples 1000 times. For MD, we compare individual trajectories. Inference time is reported for an example protein of length 300. Best values are in bold, second-best underlined.
    }
    \resizebox{\textwidth}{!}{%
    \begin{tabular}{lcccccc}
    \toprule
     & \multicolumn{2}{c}{ATLAS test set} & \multicolumn{2}{c}{De novo proteins} & \# Params & Time [s] \\
    \cmidrule(lr){2-3} \cmidrule(lr){4-5}
     & Per-target $r$ ($\uparrow$) & MAE [\AA] ($\downarrow$) & Per-target $r$ ($\uparrow$) & MAE [\AA] ($\downarrow$) & &  \\
    \midrule
    MD (Ground Truth) & 0.88 (0.01) & 0.47 (0.00) & 0.84 (0.02) & 0.19 (0.01) & -- & $\mathcal{O}(10,000)$ \\
    \midrule
    FlexPert$^{\dagger}$ & 0.83 (0.01) & 0.80 (0.02) & 0.57 (0.04) & 0.67 (0.02) & 1.2B & 0.4 \\
    Pegasus$^{\dagger}$ & 0.75 (0.01) & 0.82 (0.02) & 0.65 (0.03) & 0.74 (0.01) & 11M & 2.5 \\ 
    BackFlip & \underline{0.84} (0.01) & \underline{0.62} (0.00) & \underline{0.74} (0.01) & \underline{0.55} (0.01) & \textbf{321K} & $\leq$ \textbf{0.02} \\ 
    \midrule
    \textbf{BackFlip-2}$^{\star}$ & \textbf{0.87} (0.01) & \textbf{0.58} (0.00) & \textbf{0.78} (0.01) & \textbf{0.34} (0.00) & \underline{965K} & $\leq$ \textbf{0.02} \\
    \bottomrule
\end{tabular}
    }%
    \label{tab:rmsf_performance}
    \footnotesize{$^{\dagger}$ pLM embeddings $^{\star}$ One-hot sequence encoding}
\end{table*}

\begin{table*}
    \centering
    \caption{BackFlip-2 outperforms DynaProt and is competitive with large ensemble generation models. We report the performance on the task of predicting per-residue covariance and pairwise residue couplings on the ATLAS test set (AlphaFlow split). All metrics are reported as a median over test set proteins, as in AlphaFlow, with deviation computed by bootstrapping MD trajectories 1000 times. For ensemble generation models, the total inference time required to generate 200 conformations (as in ~\cite{alphaflow}) for a protein of 300 residues is reported.}
    \vspace{0.2cm}
    \resizebox{\textwidth}{!}{%
    \begin{tabular}{@{}lcccccccc@{}}
\toprule
\multirow{2}{*}{Method}
& \multicolumn{2}{c}{Variance contrib. of}
& \multicolumn{2}{c}{RMSF}
& \multicolumn{2}{c}{DCCM}
& \multirow{2}{*}{\# Params}
& \multirow{2}{*}{Time [s]} \\
\cmidrule(lr){2-3}
\cmidrule(lr){4-5}
\cmidrule(lr){6-7}
& RMWD ($\downarrow$)
& KL ($\downarrow$)
& $r$ ($\uparrow$)
& MAE ($\downarrow$)
& $r$ ($\uparrow$)
& MAE ($\downarrow$)
& 
&  \\
\midrule

MD (Gr. Tr.)
& 0.78 (0.01)
& 0.57 (0.00)
& 0.91 (0.00)
& 0.34 (0.00)
& 0.85 (0.00)
& 0.14 (0.00)
& -
& $\mathcal{O}(10{,}000)$ \\
\midrule

NMA (ANM)
& 1.46 (0.02)
& 4.55 (0.07)
& 0.79 (0.00)
& 1.03 (0.02)
& 0.78 (0.00)
& 0.20 (0.0)
& 31M
& $\sim 2$ \\

BioEmu
& 1.99 (0.03)
& 2.66 (0.05)
& 0.84 (0.00)
& 1.40 (0.05)
& 0.80 (0.00)
& 0.19 (0.00)
& 31M
& $\sim 380$ \\

AFMD-T
& \textbf{0.84} (0.01)
& \textbf{0.60} (0.00)
& \textbf{0.93} (0.00)
& \textbf{0.25} (0.00)
& \textbf{0.89} (0.00)
& \textbf{0.12} (0.00)
& 95M
& $\sim 8200$ \\

BBFlow
& 0.93 (0.01)
& \underline{0.65} (0.01)
& \textbf{0.93} (0.00)
& 0.42 (0.00)
& \underline{0.86} (0.00)
& \underline{0.16} (0.00)
& 18.2M
& $\sim 160$ \\
\midrule

DynaProt
& 1.18
& 0.91
& 0.87
& -
& 0.66
& -
& \underline{955K / 2.86M$^{\star}$}
& \textbf{0.02} \\

\textbf{BackFlip-2}
& \underline{0.88} (0.00)
& 0.67 (0.01)
& 0.89 (0.00)
& \underline{0.35} (0.00)
& 0.80 (0.00)
& \underline{0.16} (0.00)
& \textbf{965K}
& $\leq \textbf{0.02}$ \\
\bottomrule
\end{tabular}



    
    


    
    \label{tab:covar_performance}
    }%
\footnotesize{ $^{\star}$ Separately trained models for per-residue covariance / pairwise couplings.}
\vspace{-0.3cm}
\end{table*}

\paragraph{Predicting per-residue covariances and pairwise couplings}
We next compare BackFlip-2 to DynaProt, which is trained to predict the same observables, and find that our model performs substantially better across all evaluated metrics. In particular, our model achieves higher accuracy for the predicted per-residue covariances (RMWD Var. and KL Var.), yields more accurate reconstructed RMSF, and most notably captures dynamic residue couplings far better than DynaProt while having almost 3 times less trainable parameters (Tab.~\ref{tab:covar_performance}). The improvement in metrics is well above the finite-sampling noise of the reference MD trajectories (see App.~\ref{sec:app_convergence}). At the same time, BackFlip-2 remains competitive with the much larger ensemble generation models AlphaFlow and BBFlow on per-residue covariance metrics, but underperforms on predicting residue couplings. However, our model is over 5 orders of magnitude faster than AlphaFlow and 4 orders of magnitude faster than BBFlow. Examples of predicted pairwise residue couplings are illustrated in Fig. \ref{fig:exp_rmsf_couplings}.

\subsection{Evaluation of anisotropy of the predicted covariances}
\label{sec:anisotropy}

\begin{wraptable}{r}{0.5\textwidth}
    \vspace{-0.6cm}
    \centering
    \caption{Evaluation of covariance directionality on the ATLAS test set, measured
    by volume overlap of the induced ellipsoids. IoU: Intersection over Union. Best
    values in bold, second-best underlined.}
    \scriptsize
    \resizebox{0.5\textwidth}{!}{%
    \begin{tabular}{lcccc}
\toprule
Model
& Equiv.
& Ensemble gen.
& IoU
& Dice \\
&
&
& ($\uparrow$)
& ($\uparrow$) \\
\midrule
Ours
& \checkmark
& $\times$
& \textbf{0.49}
& \textbf{0.64} \\
Ours
& $\times$
& $\times$
& 0.42
& 0.58 \\
\midrule
BioEmu
& -
& \checkmark
& 0.25
& 0.38 \\
AFMD-T
& -
& \checkmark
& \underline{0.48}
& \underline{0.63} \\
BBFlow
& -
& \checkmark
& 0.47
& 0.62 \\
\bottomrule
\end{tabular}
    }
    \label{tab:model_ablation_iou}
    \vspace{-0.6cm}
\end{wraptable}

For each predicted covariance matrix $\widehat{\Sigma}^{(ii)}$, the eigenvectors define the principal directions of fluctuation and the corresponding eigenvalues determine the magnitude of fluctuations along these directions. Since the directionality of residue motions is of particular interest for describing functional protein dynamics (see Sec.~\ref{sec:introduction}), we explicitly evaluate the shape of the predicted covariances. To this end, we construct ellipsoids from the eigendecomposition of the predicted and MD covariance matrices and quantify their volume overlap using IoU and Dice metrics, as described in App.~\ref{sec:app_metrics}. We find that enforcing $\mathrm{SE}(3)$ equivariance in the covariance prediction head substantially improves ellipsoid volume overlap with MD compared to an $\mathrm{SE}(3)$-invariant model variant (Tab. \ref{tab:model_ablation_iou}). Crucially, our proposed equivariant model recovers anisotropic ellipsoids that match MD (Fig.~\ref{fig:exp_ellipsoids}), whereas the invariant parametrization yields nearly isotropic predicted covariances with spherical shape. Notably, our model also slightly outperforms expensive ensemble generative models AFMD-T and BBFlow at capturing the anisotropy of fluctuations.

\begin{figure*}
    \centering
    \includegraphics[width=.9\textwidth]{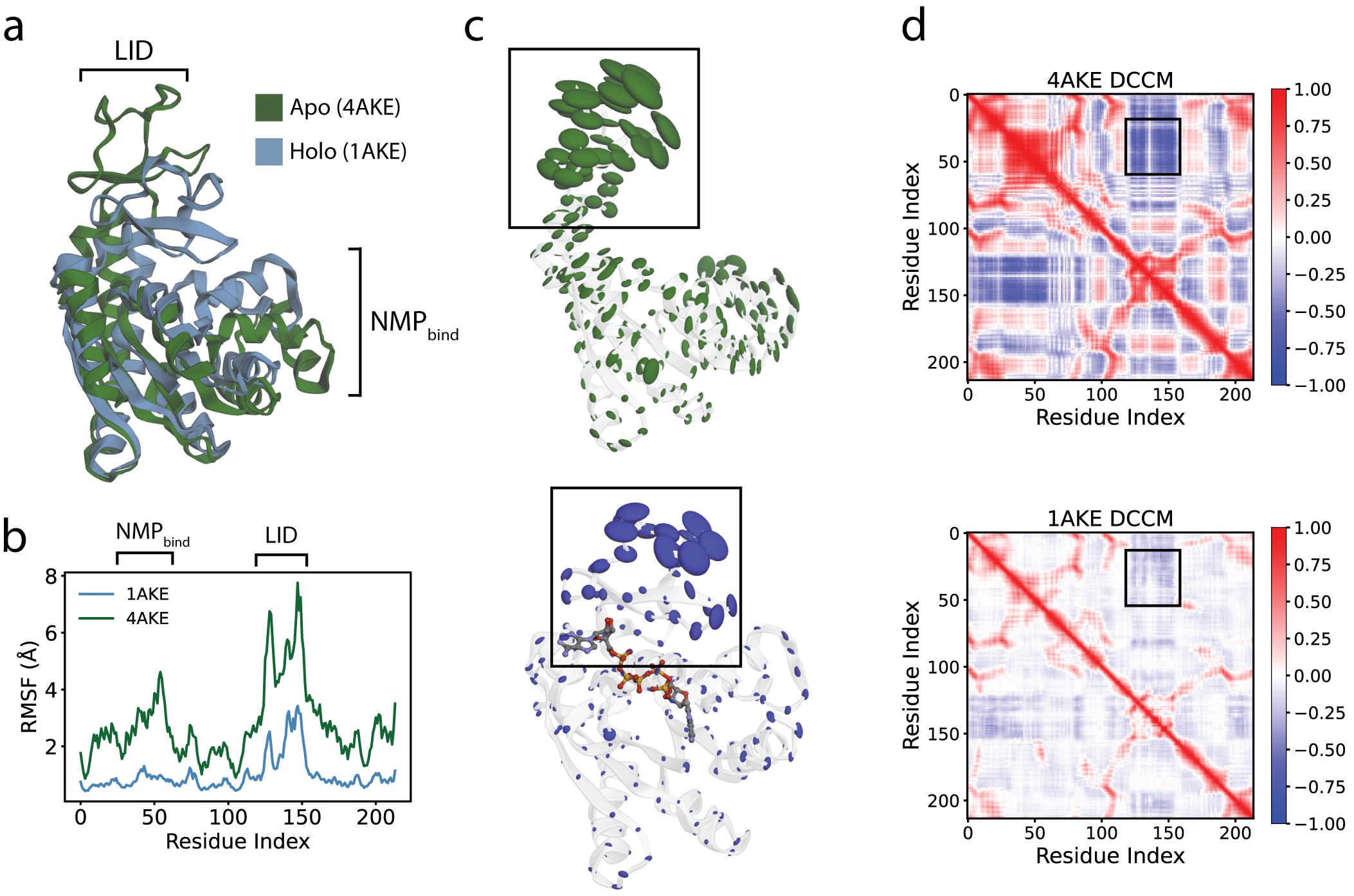}
    \caption{BackFlip-2 predictions for Adenylate Kinase (AdK).
        (a) The apo (4AKE) $\rightarrow$ holo (1AKE) major conformational transition ($\sim$8\,\AA) closes the LID and NMPbind domains over the active site. (b) Both domains are predicted to be more flexible in the apo state, with (c) highly anisotropic motions along the closure-opening axis and (d) strongly negative pairwise couplings between LID and NMPbind residues, capturing their concurrent movement towards each other.}
    \label{fig:exp_adk_main}
\end{figure*}

\subsection{Evaluation on functional protein motions and NMR ensembles}
\label{sec:bio_examples}

We predicted covariance ellipsoids and pairwise couplings for adenylate kinase (AdK), which closes its LID and NMPbind domains over the active site in the $\sim$\,8\,\AA{} apo (4AKE) to holo (1AKE) transition. In the apo state the model predicts substantially higher flexibility, especially in the LID and NMPbind domains, with the highly anisotropic predicted movement oriented toward the active site, and substantially stronger anti-correlated NMPbind-LID coupling (i.e. movement towards each other) compared with the holo state (Fig.~\ref{fig:exp_adk_main}). This state-specific differentiation is consistent with the established mechanism of domain closure~\cite{muller1996adenylate, sinev1996domain, ping2013molecular} and rigidification of AdK~\cite{sinev1996domain, shapiro2000backbone}, while the persistence of the negative LID--NMPbind coupling in the compact, rigidified holo state suggests that the apo-holo interconversion proceeds through concerted movement of the two domains. We provide additional results for dihydrofolate reductase (DHFR), where the model predicts functionally relevant catalytic Met20 loop behavior despite a $\sim$ 0.8\,\AA{} RMSD between open and closed conformations, in App.~\ref{sec:app_dhfr}. 
Furthermore, without any training on NMR data, the model very closely reproduces the structural heterogeneity and coupling patterns of NMR ensembles and outperforms direct flexibility-prediction baselines, further suggesting transfer beyond the MD settings (App.~\ref{sec:app_nmr_ensembles} and Tab. \ref{tab:app_nmr_performance}).

\subsection{Extension to the mdCATH dataset}

We additionally evaluate ATLAS-trained BackFlip-2 on the mdCATH dataset to assess generalization across simulation settings, as in FlexPert~\cite{flexpert}. 
We find that the ATLAS-trained model transfers well to the longer 500 ns timescale and higher-temperature mdCATH setting (see Sec. \ref{sec:related_work}), particularly on RMSF and pairwise residue couplings, whereas FlexPert and the predecessor BackFlip show drop in performance compared to their results on ATLAS (Tab.~\ref{tab:app_mdcath_performance}). In addition, we train BackFlip-2 on a joint ATLAS/mdCATH dataset (App.~\ref{sec:app_ATLAS_mdCATH_dataset}). Joint training yields only modest gains on mdCATH, with slight improvements for per-residue covariance predictions. If evaluated on the ATLAS test set, incorporating mdCATH data into training slightly degrades performance (Tab. \ref{tab:app_joint_trained_on_atlas}). Overall, the ATLAS-only model remains robust and readily applicable across different simulation regimes.

\subsection{Ablations}
\label{sec:ablation}

\paragraph{Ablating model components}

We ablate the use of $\mathrm{SE}(3)$ equivariance and the SPSD parametrization proposed in Sec.~\ref{sec:dynamics_prediction_module} and report the results in Tab.~\ref{tab:app_model_ablation}. We find that replacing equivariance by invariance in the per-residue covariance head leads to a substantial decrease in the performance, as can also be seen in Fig. \ref{fig:exp_ellipsoids}. Further, the simple SPDS parametrization proposed in Eq.~\ref{eq:local_cov} performs slightly better than the Cholesky factorization approach introduced in DynaProt~\cite{dynaprot} (entry b).

\paragraph{Learning curve}

To investigate how performance of the proposed model scales with the amount of training data, we report performance on the ATLAS test set as a function of the training dataset size (Fig.~\ref{fig:learning_curve}). With as little as 10\% of the training data, BackFlip-2 matches the RMSF prediction performance of Pegasus, FlexPert and the predecessor BackFlip, and reaches the performance of DynaProt with approximately 50\% of the data. For per-residue covariances, our model outperforms DynaProt using only 10\% of the training data, showcasing the importance of $\mathrm{SE}(3)$ equivariance as inductive bias for this task.

\paragraph{Noised or AlphaFold-predicted structures as input}
We further test sensitivity of the model to the quality of the input backbone by perturbing the equilibrium structures of ATLAS proteins with Gaussian noise and by replacing them with AlphaFold2 (AF2) predictions (Tab.~\ref{tab:app_noised_inputs}). Performance is only slightly affected in both settings, even though the AF2 structures deviate from the equilibrium states by a median global RMSD of $\sim$ 2.6~\AA. This suggests the model is readily applicable to predicted or moderately perturbed backbones.


\subsection{Limitations}
\label{sec:limitations}
Compared to the large ensemble generative models trained on the ATLAS dataset, BackFlip-2 is slightly less accurate, however, it is substantially faster at inference and outperforms the more comparable direct flexibility prediction baselines. As described in Sec.~\ref{sec:introduction}, the approach is fundamentally limited to predicting anisotropic motions and couplings occuring on the 300ns and 500ns MD timescale covered in the ATLAS and mdCATH datasets. It thus primarily captures sub-microsecond transitions and local degrees of freedom and cannot be expected to accurately describe slow, large-scale conformational changes, similar to all considered baselines except for BioEmu. This is because (i) the training data does not contain longer MD trajectories and (ii) the dynamical descriptors from Sec.~\ref{sec:targets_description} are suitable for fluctuations that are approximately Gaussian, as we also discuss in App. \ref{sec:app_limitations}, which is not the case for typical millisecond transitions between distant conformations.
\section{Conclusions}
\label{sec:conclusion}

We introduced BackFlip-2, an SE(3)-equivariant neural network for predicting directional flexibility and pairwise residue couplings from the equilibrium structure of a protein. We demonstrated that our method achieves state-of-the-art performance in terms of accuracy and inference time among flexibility prediction models on the established ATLAS benchmark dataset. Our method is either slightly less accurate or on-par with substantially larger ensemble generation models while being orders of magnitude faster. In ablations we show that SE(3) equivariance is a well-suited inductive bias for learning anisotropic covariances, whereas invariant models tend to predict spherical covariances and hence underperform.
Correctly predicting these directional degrees of freedom is especially important in practice since anisotropic residue motions describe functional protein rearrangements, including hinge motion, allosteric communication, and molecular recognition~\cite{zhou2014aligning, teilum2009functional, guo2016protein}. We therefore expect our model to serve as a fast and accurate tool for characterizing anisotropic functional dynamics across large de novo or natural protein datasets such as the AlphaFoldDB \cite{varadi2022alphafold}, especially in cases where ensembles are not necessarily required - for instance to identify flexible functional sites, map residue coupling patterns, or screen for putative allosteric networks across protein families. 
\newpage
\bibliographystyle{unsrtnat} 
\bibliography{bibliography_bbflow}

@article{flexpert,
  title={Learning to engineer protein flexibility},
  author={Kouba, Petr and Planas-Iglesias, Joan and Damborsky, Jiri and Sedlar, Jiri and Mazurenko, Stanislav and Sivic, Josef},
  journal={arXiv preprint arXiv:2412.18275},
  year={2024}
}

@incollection{hayward2008normal,
  title={Normal modes and essential dynamics},
  author={Hayward, Steven and De Groot, Bert L},
  booktitle={Molecular Modeling of Proteins},
  pages={89--106},
  year={2008},
  publisher={Springer}
}

@incollection{hinsen2005normal,
  title={Normal mode theory and harmonic potential approximations},
  author={Hinsen, Konrad},
  booktitle={Normal Mode Analysis},
  pages={25--40},
  year={2005},
  publisher={Chapman and Hall/CRC}
}

@article{vander2024atlas,
  title={ATLAS: protein flexibility description from atomistic molecular dynamics simulations},
  author={Vander Meersche, Yann and Cretin, Gabriel and Gheeraert, Aria and Gelly, Jean-Christophe and Galochkina, Tatiana},
  journal={Nucleic acids research},
  volume={52},
  number={D1},
  pages={D384--D392},
  year={2024},
  publisher={Oxford University Press}
}

@article{flips,
  title={Flexibility-conditioned protein structure design with flow matching},
  author={Viliuga, Vsevolod and Seute, Leif and Wolf, Nicolas and Wagner, Simon and Elofsson, Arne and St{\"u}hmer, Jan and Gr{\"a}ter, Frauke},
  journal={arXiv preprint arXiv:2508.18211},
  year={2025}
}

@article{mdcath,
  title={mdCATH: A large-scale md dataset for data-driven computational biophysics},
  author={Mirarchi, Antonio and Giorgino, Toni and De Fabritiis, Gianni},
  journal={Scientific Data},
  volume={11},
  number={1},
  pages={1299},
  year={2024},
  publisher={Nature Publishing Group UK London}
}

@article{orengo1999cath,
  title={The CATH Database provides insights into protein structure/function relationships},
  author={Orengo, Christine A and Pearl, Frances MG and Bray, James E and Todd, Annabel E and Martin, AC and Lo Conte, L and Thornton, Janet M},
  journal={Nucleic acids research},
  volume={27},
  number={1},
  pages={275--279},
  year={1999},
  publisher={Oxford University Press}
}

@article{vander2025pegasus,
  title={PEGASUS: Prediction of MD-derived protein flexibility from sequence},
  author={Vander Meersche, Yann and Duval, Gabriel and Cretin, Gabriel and Gheeraert, Aria and Gelly, Jean-christophe and Galochkina, Tatiana},
  journal={Protein Science},
  volume={34},
  number={8},
  pages={e70221},
  year={2025},
  publisher={Wiley Online Library}
}

@article{alphaflow,
  title={AlphaFold meets flow matching for generating protein ensembles},
  author={Jing, Bowen and Berger, Bonnie and Jaakkola, Tommi},
  journal={arXiv preprint arXiv:2402.04845},
  year={2024}
}

@article{alphafold2,
  title={Highly accurate protein structure prediction with AlphaFold},
  author={Jumper, John and Evans, Richard and Pritzel, Alexander and Green, Tim and Figurnov, Michael and Ronneberger, Olaf and Tunyasuvunakool, Kathryn and Bates, Russ and {\v{Z}}{\'\i}dek, Augustin and Potapenko, Anna and others},
  journal={nature},
  volume={596},
  number={7873},
  pages={583--589},
  year={2021},
  publisher={Nature Publishing Group UK London}
}

@article{bioemu,
  title={Scalable emulation of protein equilibrium ensembles with generative deep learning},
  author={Lewis, Sarah and Hempel, Tim and Jim{\'e}nez-Luna, Jos{\'e} and Gastegger, Michael and Xie, Yu and Foong, Andrew YK and Satorras, Victor Garc{\'\i}a and Abdin, Osama and Veeling, Bastiaan S and Zaporozhets, Iryna and others},
  journal={Science},
  volume={389},
  number={6761},
  pages={eadv9817},
  year={2025},
  publisher={American Association for the Advancement of Science}
}

@article{bbflow,
  title={Learning conformational ensembles of proteins based on backbone geometry},
  author={Wolf, Nicolas and Seute, Leif and Viliuga, Vsevolod and Wagner, Simon and St{\"u}hmer, Jan and Gr{\"a}ter, Frauke},
  journal={arXiv preprint arXiv:2503.05738},
  year={2025}
}

@article{dynaprot,
  title={Learning residue level protein dynamics with multiscale Gaussians},
  author={Bafna, Mihir and Jing, Bowen and Berger, Bonnie},
  journal={ArXiv},
  pages={arXiv--2509},
  year={2025}
}

@inproceedings{cmd,
  title={Correlation matrix distance, a meaningful measure for evaluation of non-stationary MIMO channels},
  author={Herdin, Markus and Czink, Nicolai and Ozcelik, H{\"u}seyin and Bonek, Ernst},
  booktitle={2005 IEEE 61st vehicular technology conference},
  volume={1},
  pages={136--140},
  year={2005},
  organization={IEEE}
}

@article{foldseek,
  title={Fast and accurate protein structure search with Foldseek},
  author={Van Kempen, Michel and Kim, Stephanie S and Tumescheit, Charlotte and Mirdita, Milot and Lee, Jeongjae and Gilchrist, Cameron LM and S{\"o}ding, Johannes and Steinegger, Martin},
  journal={Nature biotechnology},
  volume={42},
  number={2},
  pages={243--246},
  year={2024},
  publisher={Nature Publishing Group US New York}
}

@article{huang2025novo,
  title={De Novo Design, Directed Evolution and Computational Study of Heme-Binding Helical Bundle Protein Catalysts for Biocatalytic Enantioselective Ge--H Insertion},
  author={Huang, Wei and Adornato, Gessica M and Horst, Maggie and Alturaifi, Turki M and Hou, Kaipeng and Liu, Peng and DeGrado, William F and Yang, Yang},
  journal={Journal of the American Chemical Society},
  volume={147},
  number={44},
  pages={40869--40878},
  year={2025},
  publisher={ACS Publications}
}

@article{khmelinskaia2025local,
  title={Local structural flexibility drives oligomorphism in computationally designed protein assemblies},
  author={Khmelinskaia, Alena and Bethel, Neville P and Fatehi, Farzad and Mallik, Bhoomika Basu and Antanasijevic, Aleksandar and Borst, Andrew J and Lai, Szu-Hsueh and Chim, Ho Yeung and Wang, Jing Yang ‘John’ and Miranda, Marcos C and others},
  journal={Nature Structural \& Molecular Biology},
  volume={32},
  number={6},
  pages={1050--1060},
  year={2025},
  publisher={Nature Publishing Group US New York}
}

@article{du2025conformational,
  title={Conformational ensembles reveal the origins of serine protease catalysis},
  author={Du, Siyuan and Kretsch, Rachael C and Parres-Gold, Jacob and Pieri, Elisa and Cruzeiro, Vin{\'\i}cius Wilian D and Zhu, Mingning and Pinney, Margaux M and Yabukarski, Filip and Schwans, Jason P and Mart{\'\i}nez, Todd J and others},
  journal={Science},
  volume={387},
  number={6735},
  pages={eado5068},
  year={2025},
  publisher={American Association for the Advancement of Science}
}

@article{reiss2020allosteric,
  title={Allosteric control of enzyme activity: from ancient origins to recent gene-editing technologies},
  author={Reiss, Krystle and Batista, Victor S and Loria, J Patrick},
  journal={Biochemistry},
  volume={59},
  number={18},
  pages={1711--1712},
  year={2020},
  publisher={ACS Publications}
}

@article{tousignant2004protein,
  title={Protein motions promote catalysis},
  author={Tousignant, Audrey and Pelletier, Joelle N},
  journal={Chemistry \& biology},
  volume={11},
  number={8},
  pages={1037--1042},
  year={2004},
  publisher={Elsevier}
}

@article{schwartz2023protein,
  title={Protein dynamics and enzymatic catalysis},
  author={Schwartz, Steven D},
  journal={The Journal of Physical Chemistry B},
  volume={127},
  number={12},
  pages={2649--2660},
  year={2023},
  publisher={ACS Publications}
}

@article{nam2023protein,
  title={Protein dynamics: The future is bright and complicated!},
  author={Nam, Kwangho and Wolf-Watz, Magnus},
  journal={Structural Dynamics},
  volume={10},
  number={1},
  year={2023},
  publisher={AIP Publishing}
}

@article{kurkcuoglu2012coupling,
  title={Coupling between catalytic loop motions and enzyme global dynamics},
  author={Kurkcuoglu, Zeynep and Bakan, Ahmet and Kocaman, Duygu and Bahar, Ivet and Doruker, Pemra},
  year={2012},
  publisher={Public Library of Science San Francisco, USA}
}

@article{spyrakis2011protein,
  title={Protein flexibility and ligand recognition: challenges for molecular modeling},
  author={Spyrakis, Francesca and BidonChanal, Axel and Barril, Xavier and Javier Luque, F},
  journal={Current topics in medicinal chemistry},
  volume={11},
  number={2},
  pages={192--210},
  year={2011},
  publisher={Bentham Science Publishers}
}

@article{csermely2010induced,
  title={Induced fit, conformational selection and independent dynamic segments: an extended view of binding events},
  author={Csermely, Peter and Palotai, Robin and Nussinov, Ruth},
  journal={Trends in biochemical sciences},
  volume={35},
  number={10},
  pages={539--546},
  year={2010},
  publisher={Elsevier}
}

@book{voetvoet,
  title={Biochemistry},
  author={Voet, Donald and Voet, Judith G},
  year={2010},
  publisher={John Wiley \& Sons}
}

@article{brooks1983harmonic,
  title={Harmonic dynamics of proteins: normal modes and fluctuations in bovine pancreatic trypsin inhibitor.},
  author={Brooks, Bernard and Karplus, Martin},
  journal={Proceedings of the National Academy of Sciences},
  volume={80},
  number={21},
  pages={6571--6575},
  year={1983}
}

@inproceedings{ho2020ddpm,
  title={Denoising Diffusion Probabilistic Models},
  author={Ho, Jonathan and Jain, Ajay and Abbeel, Pieter},
  booktitle={Advances in Neural Information Processing Systems},
  volume={33},
  pages={6840--6851},
  year={2020}
}

@article{lipman2022flow,
  title   = {Flow matching for generative modeling},
  author  = {Lipman, Yaron and Chen, Ricky TQ and Ben-Hamu, Heli and Nickel, Maximilian and Le, Matt},
  journal = {International Conference on Learning Representations},
  year    = {2023}
}

@inproceedings{albergo2022building,
  title={Building Normalizing Flows with Stochastic Interpolants},
  author={Albergo, M. S. and Vanden-Eijnden, E.},
  booktitle={The Eleventh International Conference on Learning Representations},
  year={2022}
}

@inproceedings{wang2024proteinconfdiff,
  title={Protein Conformation Generation via Force-Guided SE (3) Diffusion Models},
  author={Wang, Yan and Wang, Lihao and Shen, Yuning and Wang, Yiqun and Yuan, Huizhuo and Wu, Yue and Gu, Quanquan},
  booktitle={Forty-first International Conference on Machine Learning},
  year={2024}
}

@inproceedings{
jing2024mdgen,
title={Generative Modeling of Molecular Dynamics Trajectories},
author={Bowen Jing and Hannes Stark and Tommi Jaakkola and Bonnie Berger},
booktitle={The Thirty-eighth Annual Conference on Neural Information Processing Systems},
year={2024},
}

@article{zhou2014aligning,
  title={Aligning experimental and theoretical anisotropic B-factors: water models, normal-mode analysis methods, and metrics},
  author={Zhou, Lei and Liu, Qinglian},
  journal={The Journal of Physical Chemistry B},
  volume={118},
  number={15},
  pages={4069--4079},
  year={2014},
  publisher={ACS Publications}
}

@article{guo2016protein,
  title={Protein allostery and conformational dynamics},
  author={Guo, Jingjing and Zhou, Huan-Xiang},
  journal={Chemical reviews},
  volume={116},
  number={11},
  pages={6503--6515},
  year={2016},
  publisher={ACS Publications}
}

@article{teilum2009functional,
  title={Functional aspects of protein flexibility},
  author={Teilum, Kaare and Olsen, Johan G and Kragelund, Birthe B},
  journal={Cellular and Molecular Life Sciences},
  volume={66},
  number={14},
  pages={2231--2247},
  year={2009},
  publisher={Springer}
}

@article{wayment-steele2024predicting,
  author    = {Hannah K. Wayment-Steele and Adedolapo Ojoawo and Renee Otten and Julia M. Apitz and Warintra Pitsawong and Marc H{\"o}mberger and Sergey Ovchinnikov and Lucy Colwell and Dorothee Kern},
  title = {Predicting multiple conformations via sequence clustering and AlphaFold2},
  journal = {Nature},
  volume = {625},
  pages = {832--839},
  year = {2024},
}

@article{watson2023novo,
  author = {Watson, Joseph L. and Juergens, David and Bennett, Nathaniel R. and Trippe, Brian L. and Yim, Jason and Eisenach, Helen E. and Ahern, Woody and Borst, Andrew J. and Ragotte, Robert J. and Milles, Lukas F. and Wicky, Basile I. M. and Hanikel, Nikita and Pellock, Samuel J. and Courbet, Alexis and Sheffler, William and Wang, Jue and Venkatesh, Preetham and Sappington, Isaac and Vázquez Torres, Susana and Lauko, Anna and De Bortoli, Valentin and Mathieu, Emile and Ovchinnikov, Sergey and Barzilay, Regina and Jaakkola, Tommi S. and DiMaio, Frank and Baek, Minkyung and Baker, David},
  title = {De novo design of protein structure and function with RFdiffusion},
  journal   = {Nature},
  pages     = {1--3},
  year      = {2023},
  publisher = {Nature Publishing Group UK London}
}

@article{yim2023frameflow,
  title={Fast protein backbone generation with SE (3) flow matching},
author={Jason Yim and Andrew Campbell and Andrew Y. K. Foong and Michael Gastegger and José Jiménez-Luna and Sarah Lewis and Victor Garcia Satorras and Bastiaan S. Veeling and Regina Barzilay and Tommi Jaakkola and Frank Noé},
  journal={arXiv preprint arXiv:2310.05297},
  year={2023}
}

@article{twomey2018structural,
  title={Structural mechanisms of gating in ionotropic glutamate receptors},
  author={Twomey, Edward C and Sobolevsky, Alexander I},
  journal={Biochemistry},
  volume={57},
  number={3},
  pages={267--276},
  year={2018},
  publisher={ACS Publications}
}

@article{galenkamp2024allostery,
  title={Allostery can convert binding free energies into concerted domain motions in enzymes},
  author={Galenkamp, Nicole St{\'e}phanie and Zernia, Sarah and Van Oppen, Yulan B and van den Noort, Marco and Milias-Argeitis, Andreas and Maglia, Giovanni},
  journal={Nature Communications},
  volume={15},
  number={1},
  pages={10109},
  year={2024},
  publisher={Nature Publishing Group UK London}
}

@article{muller1996adenylate,
  title={Adenylate kinase motions during catalysis: an energetic counterweight balancing substrate binding},
  author={M{\"u}ller, Christoph W and Schlauderer, Gerwald J and Reinstein, Jochen and Schulz, Georg E},
  journal={Structure},
  volume={4},
  number={2},
  pages={147--156},
  year={1996},
  publisher={Elsevier}
}

@article{sinev1996domain,
  title={Domain closure in adenylate kinase},
  author={Sinev, Michael A and Sineva, Elena V and Ittah, Varda and Haas, Elisha},
  journal={Biochemistry},
  volume={35},
  number={20},
  pages={6425--6437},
  year={1996},
  publisher={ACS Publications}
}

@article{ping2013molecular,
  title={Molecular dynamics studies on the conformational transitions of adenylate kinase: a computational evidence for the conformational selection mechanism},
  author={Ping, Jie and Hao, Pei and Li, Yi-Xue and Wang, Jing-Fang},
  journal={BioMed Research International},
  volume={2013},
  number={1},
  pages={628536},
  year={2013},
  publisher={Wiley Online Library}
}

@article{shapiro2000backbone,
  title={Backbone dynamics of Escherichia coli adenylate kinase at the extreme stages of the catalytic cycle studied by 15N NMR relaxation},
  author={Shapiro, Yury E and Sinev, Michael A and Sineva, Elena V and Tugarinov, Vitali and Meirovitch, Eva},
  journal={Biochemistry},
  volume={39},
  number={22},
  pages={6634--6644},
  year={2000},
  publisher={ACS Publications}
}

@article{sawaya1997loop,
  title={Loop and subdomain movements in the mechanism of Escherichia coli dihydrofolate reductase: crystallographic evidence},
  author={Sawaya, Michael R and Kraut, Joseph},
  journal={Biochemistry},
  volume={36},
  number={3},
  pages={586--603},
  year={1997},
  publisher={ACS Publications}
}

@article{osborne2001backbone,
  title={Backbone dynamics in dihydrofolate reductase complexes: role of loop flexibility in the catalytic mechanism},
  author={Osborne, Michael J and Schnell, Jason and Benkovic, Stephen J and Dyson, H Jane and Wright, Peter E},
  journal={Biochemistry},
  volume={40},
  number={33},
  pages={9846--9859},
  year={2001},
  publisher={ACS Publications}
}

@article{lindorff2011fast,
  title={How fast-folding proteins fold},
  author={Lindorff-Larsen, Kresten and Piana, Stefano and Dror, Ron O and Shaw, David E},
  journal={Science},
  volume={334},
  number={6055},
  pages={517--520},
  year={2011},
  publisher={American Association for the Advancement of Science}
}

@article{
diepeveen_2024,
author = {Willem Diepeveen  and Carlos Esteve-Yagüe  and Jan Lellmann  and Ozan Öktem  and Carola-Bibiane Schönlieb },
title = {Riemannian geometry for efficient analysis of protein dynamics data},
journal = {Proceedings of the National Academy of Sciences},
volume = {121},
number = {33},
pages = {e2318951121},
year = {2024},
}

@article{varadi2022alphafold,
  title={AlphaFold Protein Structure Database: massively expanding the structural coverage of protein-sequence space with high-accuracy models},
  author={Varadi, Mihaly and Anyango, Stephen and Deshpande, Mandar and Nair, Sreenath and Natassia, Cindy and Yordanova, Galabina and Yuan, David and Stroe, Oana and Wood, Gemma and Laydon, Agata and others},
  journal={Nucleic acids research},
  volume={50},
  number={D1},
  pages={D439--D444},
  year={2022},
  publisher={Oxford University Press}
}

\appendix

\onecolumn
\section{Appendix}

\setcounter{figure}{0}
\renewcommand\thefigure{\thesection.\arabic{figure}}
\setcounter{table}{0}
\renewcommand{\thetable}{\thesection.\arabic{table}}
\renewcommand*{\theHtable}{\thetable}
\renewcommand*{\theHfigure}{\thefigure}


\subsection{Comparison to Cholesky parametrization}
\label{sec:app_cholesky_comparison}
DynaProt~\cite{dynaprot} parametrizes the covariance head via a Cholesky factorization, predicting only the six elements of a lower-triangular factor $L$ with a softplus-constrained positive diagonal, and reconstructing the covariance as $LL^\top$.
The softplus constraint ensures that all eigenvalues of $LL^\top$ are strictly positive, enforcing SPD by construction.

Our $AA^\top$ parametrization instead allows $A \in \mathbb{R}^{3 \times 3}$ to be unconstrained, so the predicted covariance can be rank-deficient.
We argue this is the correct choice for two reasons.
First, it is physically well-motivated: a zero eigenvalue in some direction means the residue fluctuates with zero amplitude along that direction, i.e.\ it is perfectly rigid in that direction -- a valid configuration for buried or geometrically constrained residues.
Forcing strict positive definiteness via a softplus constraint introduces a systematic bias against such predictions, even when they are correct.
Second, strict positive definiteness is not required for training: where the matrix logarithm must be evaluated, a small $\varepsilon I$ shift is sufficient.
For any SPSD matrix $\Sigma$, the regularized matrix $\Sigma_\varepsilon = \Sigma + \varepsilon I$ with $\varepsilon > 0$ is SPD, since
\begin{align}
    v^\top \Sigma_\varepsilon\, v
    &= v^\top \!\left(\Sigma + \varepsilon I\right) v \notag \\
    &= v^\top \Sigma\, v + \varepsilon\, v^\top I\, v \notag \\
    &= v^\top \Sigma\, v + \varepsilon\, \|v\|^2 \notag \\
    &\geq 0 + \varepsilon\, \|v\|^2 \notag\\
    &= \varepsilon\, \|v\|^2 > 0,
\end{align}
where the inequality uses that $\Sigma$ is SPSD, i.e.\ $v^\top \Sigma\, v \geq 0$ for all $v$.
Since $v^\top \Sigma_\varepsilon\, v > 0$ for all $v \neq 0$, the matrix $\Sigma_\varepsilon$ is SPD.

\subsection{Metrics for evaluating per-residue covariances}
\label{sec:app_metrics}

\paragraph{RMWD Variance}

RMWD calculates the 2-Wasserstein distance between the predicted and reference distributions under the assumption that both distributions are Gaussian. Since our model predicts only per-residue covariance matrices and not mean coordinates, we center both distributions at the same equilibrium structure, analogous to DynaProt~\cite{dynaprot}, so that the mean contribution vanishes. We therefore report only the variance term:
\begin{equation}
\mathrm{RMWD}_{\mathrm{var}}
\left(
\widehat{\Sigma}, \Sigma
\right)
=
\sqrt{
\frac{1}{N}
\sum_{i=1}^{N}
\operatorname{Tr}
\left(
\widehat{\Sigma}^{(ii)}
+
\Sigma^{(ii)}
-
2
\left(
\widehat{\Sigma}^{(ii)}
\Sigma^{(ii)}
\right)^{1/2}
\right)}
\end{equation}
\paragraph{Symmetric KL Divergence Variance}
The symmetrized KL variance measures the divergence between the predicted and reference Gaussian distributions induced by their per-residue covariance blocks. Analogous to RMWD definition, we consider only the variance term. We compute the symmetric KL divergence variance by averaging the forward and reverse KL terms:
\begin{equation}
\mathrm{KL}_{\mathrm{symvar}}
\left(
\widehat{\Sigma}, \Sigma
\right)
=
\frac{1}{4N}
\sum_{i=1}^{N}
\left[
\operatorname{Tr}
\left(
\left(\Sigma^{(ii)}\right)^{-1}
\widehat{\Sigma}^{(ii)}
\right)
+
\operatorname{Tr}
\left(
\left(\widehat{\Sigma}^{(ii)}\right)^{-1}
\Sigma^{(ii)}
\right)
-
2d
\right],
\end{equation}
where \(d=3\) for Cartesian per-residue covariance blocks.

\paragraph{Volume overlap of ellipsoids induced by the per-residue covariance}

Although RMWD Variance and Symmetric KL Variance measure discrepancies between the distributions induced by the predicted and ground-truth covariances, they summarize disagreement as scalar distributional distances. However, we are interested in evaluating whether the predicted covariances recover the anisotropic residue fluctuations and the orientation of their principal axes, which are observed in MD (Fig. \ref{fig:exp_ellipsoids}). To explicitly assess this, we construct ellipsoids from the predicted and MD-derived per-residue covariance blocks and compare their volume overlap.

Concretely, for each per-residue covariance block
\(\Sigma^{(ii)} \in \mathbb{R}^{3 \times 3}\), we construct the centered
covariance ellipsoid
\begin{equation}
\mathcal{E}\!\left(\Sigma^{(ii)}\right)
=
\left\{
x \in \mathbb{R}^{3}
:
x^\top
\left(\Sigma^{(ii)}\right)^{-1}
x
\leq 1
\right\},
\end{equation}
where \(x\) is a centered Cartesian displacement vector. To obtain an intuition for the geometric relation between this ellipsoid and the covariance matrix, we note that the principal axes and semi-axis lengths of the ellipsoid can be obtained from the eigendecomposition of the covariance for a given residue $i$:
\begin{equation}
    \Sigma^{(ii)}
    =
    Q_i
    \operatorname{diag}
    \left(
        \lambda_{i1}, \lambda_{i2}, \lambda_{i3}
    \right)
    Q_i^\top,
    \qquad
    r_{ij}
    =
    \sqrt{\lambda_{ij}}.
\end{equation}
The columns of \(Q_i\) define the ellipsoid principal directions, and
\(r_{ij}\) gives the semi-axis length along the \(j\)-th principal direction.

For a unit direction \(u \in \mathbb{S}^2\), the radial extent of
\(\mathcal{E}(\Sigma^{(ii)})\) is
\begin{equation}
r_{\Sigma^{(ii)}}(u)
=
\left(
u^\top
\left(\Sigma^{(ii)}\right)^{-1}
u
\right)^{-1/2}.
\end{equation}
For predicted and ground-truth covariance blocks
\(\widehat{\Sigma}^{(ii)}\) and \(\Sigma^{(ii)}\), we then define the radial extents of their intersection $r_i^{\cap}$ and union $r_i^{\cup}$ as
\begin{equation}
r_i^{\cap}(u)
=
\min
\left(
r_{\widehat{\Sigma}^{(ii)}}(u),
r_{\Sigma^{(ii)}}(u)
\right),
\quad
r_i^{\cup}(u)
=
\max
\left(
r_{\widehat{\Sigma}^{(ii)}}(u),
r_{\Sigma^{(ii)}}(u)
\right).
\end{equation}

We then compute \textbf{IoU} (Intersection over Union) as the volume ratios of intersection and union of the ellipsoids,
\begin{equation}
\mathrm{IoU}_i
=
\frac{
\int_{\mathbb{S}^2}
\left[r_i^{\cap}(u)\right]^3
\, d\Omega
}{
\int_{\mathbb{S}^2}
\left[r_i^{\cup}(u)\right]^3
\, d\Omega
},
\end{equation}
and the \textbf{Dice} overlap as
\begin{equation}
\mathrm{Dice}_i
=
\frac{
2
\int_{\mathbb{S}^2}
\left[r_i^{\cap}(u)\right]^3
\, d\Omega
}{
\int_{\mathbb{S}^2}
r_{\widehat{\Sigma}^{(ii)}}(u)^3
\, d\Omega
+
\int_{\mathbb{S}^2}
r_{\Sigma^{(ii)}}(u)^3
\, d\Omega
}.
\end{equation}

IoU and Dice measure the overlap between the predicted and reference covariance-induced ellipsoids, hence measuring the similarity both in terms of magnitude but also in terms of shape, capturing differences in anisotropy.

\subsection{Joint ATLAS/mdCATH dataset construction}
\label{sec:app_ATLAS_mdCATH_dataset}

We construct a joint ATLAS/mdCATH dataset by assigning mdCATH domains to the existing AlphaFlow \cite{alphaflow} ATLAS train, validation, and test splits based on structural similarity. First, we cluster proteins from ATLAS and mdCATH using FoldSeek \cite{foldseek} in TMalign mode with a TM-score threshold of 0.5. We then use the original AlphaFlow ATLAS split assignment as a fixed reference. Each mdCATH domain is assigned to the train, validation, or test split if it clusters with at least one ATLAS protein from the corresponding split. Domains that do not cluster with any ATLAS protein are randomly assigned to the train, validation, and test splits in an 80/10/10 ratio. If an mdCATH domain is associated with ATLAS proteins from more than one split, we discard it to avoid structural overlap across splits. This procedure yields a final split that is approximately 80/10/10, with 5673 proteins in the training set, 531 in the validation set, and 508 in the test set. The joint ATLAS/mdCATH datasplit can be obtained at \url{https://github.com/graeter-group/backflip}.

\subsection{Details on flexibility prediction baselines}
\label{sec:app:flexibility_baselines}
Several deep learning (DL) models have been proposed to predict isotropic flexibility of protein backbones directly, most commonly quantified as per-residue Root Mean Square Fluctuation (RMSF), which measures the average positional deviation from the equilibrium state across a conformational ensemble. FlexPert \cite{flexpert} fine-tunes a pretrained protein Language Model (pLM) and fuses its embeddings with external structural features with a learnable CNN module. Pegasus \cite{vander2025pegasus} predicts RMSF from embeddings of 4 different pretrained pLMs with a CNN and does not require structure as input. BackFlip \cite{flips} instead proposes to infer RMSF with an SE(3)-invariant graph neural network using only protein backbone geometry. All these models are trained and evaluated on ATLAS.

\subsection{Prediction of flexibility and pairwise couplings in DHFR}
\label{sec:app_dhfr}

\begin{figure*}[h]
    \centering
    \includegraphics[width=.9\textwidth]{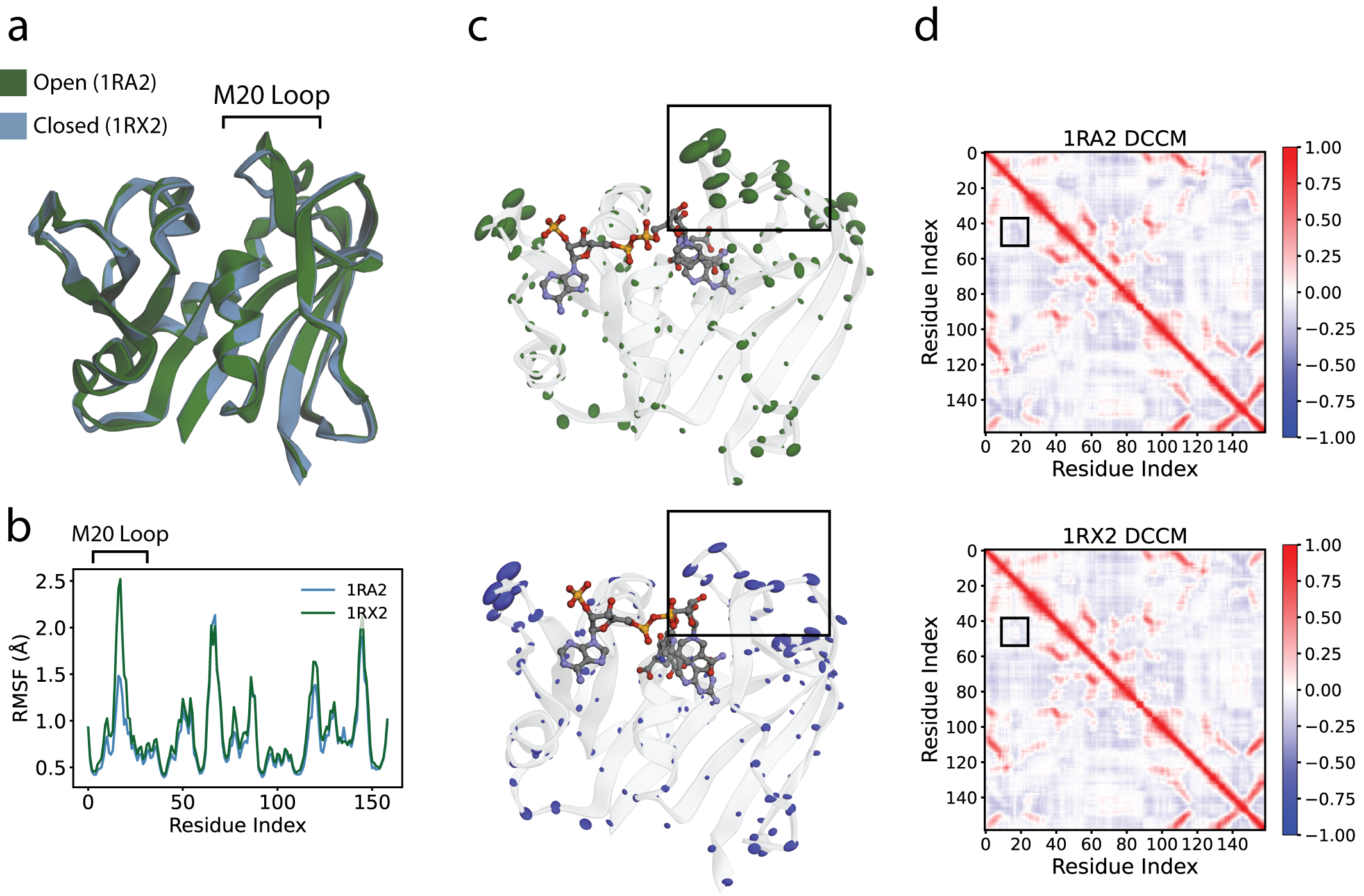}
    \caption{BackFlip-2 predictions for Dihydrofolate Reductase (DHFR). (a) DHFR undergoes a transition between the closed state (1RX2) to the occluded state proposed to occur through the open (1RA2) state, which only differ by 0.8~\AA{} in the catalytic Met20 loop. (b) BackFlip-2 predicts higher Met20-loop flexibility in the open state,
    (c) with the predicted anisotropic motion consistent with an opening-closing coordinate. (d) In the closed state, couplings between the Met20 loop and residues lining the ligand-binding site become more positive, indicating that the loop moves in concert with the binding site.}
    \label{fig:app_dhfr}
\end{figure*}

In addition to the adenylate kinase experiment (Sec. \ref{sec:bio_examples} and Fig. \ref{fig:exp_adk_main}), we evaluate the model on \textit{E.~coli} dihydrofolate reductase (DHFR), comparing the predictions made for two experimentally resolved structures with the same set of bound ligands (folate and NADP$^+$) but differing in the conformation of the catalytic Met20 loop: open (1RA2) and closed (1RX2). The open conformation has been proposed to act as an intermediate along the closed-occluded interconversion that precedes product release~\cite{sawaya1997loop}. Despite a global RMSD of only 0.8~\AA{} between the two states, the model predicts markedly higher Met20 loop flexibility in the open state, with anisotropy oriented along the closure direction (Fig.~\ref{fig:app_dhfr}). NMR relaxation experiments have reported strongly reduced loop flexibility in the closed complex, attributed to tighter packing and engagement of the Met20 loop~\cite{osborne2001backbone}. In the closed state, the model further predicts more positive couplings between the Met20 loop and residues lining the ligand-binding site, i.e. more strongly aligned motion, as could be expected if the loop moves coherently with the binding site residues.

\subsection{Performance on recapitulating the structural uncertainty of NMR ensembles}
\label{sec:app_nmr_ensembles}

We further evaluated the model on 2000 randomly selected in-solution NMR ensembles from the PDB, each with at least 10 deposited conformations. We note that deposited NMR conformations cannot be considered as a statistic, thermodynamics ensemble, they rather represent the structural uncertainty consistent with the experimental restraints, averaged over heterogeneous timescales. We therefore consider conformational variability deposited as NMR ensembles as an independent, experimentally grounded proxy for local structural heterogeneity rather than the ground truth for equilibrium populations. Despite never having been trained on NMR data, the model very closely reproduces both RMSF and the pairwise couplings (Tab.~\ref{tab:app_nmr_performance}), outperforming direct flexibility-prediction baselines.

\begin{table}[h]
    \centering
    \caption{Performance of BackFlip-2 and baselines at recapitulating structural heterogeneity of 2000 randomly selected solution NMR ensembles from the PDB. We do not report FlexPert~\cite{flexpert} performance here because inference crashes in the preprocessing pipeline of the released code. DynaProt~\cite{dynaprot} is omitted since the code was not made publicly available.}
    \vspace{0.1cm}
    \resizebox{\textwidth}{!}{%
    \begin{tabular}{lcccccc}
    \toprule
    Method 
    & RMWD Var. ($\downarrow$) 
    & KL Var. ($\downarrow$)
    & RMSF $r$ ($\uparrow$)
    & RMSF MAE ($\downarrow$)
    & DCCM $r$ ($\uparrow$)
    & DCCM MAE ($\downarrow$) \\
    \midrule
    Pegasus
    & -
    & -
    & 0.83
    & 0.93
    & -
    & - \\
    BackFlip
    & -
    & -
    & 0.90
    & 0.70
    & -
    & - \\
    \midrule
    \textbf{BackFlip-2}
    & 1.2
    & 1.9
    & \textbf{0.91}
    & \textbf{0.62}
    & 0.80
    & 0.18 \\
    \bottomrule
    \end{tabular}
    }%
    \label{tab:app_nmr_performance}
\end{table}

\newpage
\subsection{Effect of joint training}

\begin{table*}[h!]
    \centering
    \caption{Performance on 508 mdCATH proteins at 320\,K for models trained either on ATLAS alone or jointly on ATLAS/mdCATH, compared with FlexPert (not trained on mdCATH). Metrics are reported as median over all proteins in the mdCATH dataset. The description on construction of the joint ATLAS/mdCATH dataset can be found in App. \ref{sec:app_ATLAS_mdCATH_dataset}. Note that we cannot compare to DynaProt \cite{dynaprot} on the same dataset because the source code is not available.}
    \vspace{0.1cm}
    \resizebox{\textwidth}{!}{%
    \begin{tabular}{lcccccc}
    \toprule
    Method 
    & RMWD Var. ($\downarrow$) 
    & KL Var. ($\downarrow$)
    & RMSF $r$ ($\uparrow$)
    & RMSF MAE ($\downarrow$)
    & DCCM $r$ ($\uparrow$)
    & DCCM MAE ($\downarrow$) \\
    \midrule

    FlexPert 
    & - 
    & -
    & 0.70
    & 0.98
    & -
    & - \\ 

    BackFlip 
    & - 
    & -
    & 0.77
    & 0.84
    & -
    & - \\ 

    \midrule
    
    BackFlip-2-ATLAS
    & \underline{1.63}
    & \underline{2.57}
    & \underline{0.85}
    & \underline{0.69}
    & \textbf{0.81}
    & \underline{0.15} \\ 

    BackFlip-2-joint
    & \textbf{1.51}
    & \textbf{1.72}
    & \textbf{0.88}
    & \textbf{0.65}
    & \textbf{0.81}
    & \textbf{0.14} \\ 
    \bottomrule
    \end{tabular}
    }%
    \label{tab:app_mdcath_performance}
\end{table*}

\begin{table}[h!]
    \centering
    \caption{Performance on 82 ATLAS test proteins (AlphaFlow split) for models trained either on ATLAS alone or jointly on ATLAS/mdCATH. Metrics are reported as median over all proteins in the mdCATH dataset. The description on construction of the joint ATLAS/mdCATH dataset can be found in App. \ref{sec:app_ATLAS_mdCATH_dataset}}
    \vspace{0.1cm}
    \resizebox{\textwidth}{!}{%
    \begin{tabular}{lcccccc}
    \toprule
    Method 
    & RMWD Var. ($\downarrow$) 
    & KL Var. ($\downarrow$)
    & RMSF $r$ ($\uparrow$)
    & RMSF MAE ($\downarrow$)
    & DCCM $r$ ($\uparrow$)
    & DCCM MAE ($\downarrow$) \\
    \midrule

    BackFlip-2-ATLAS
    & \textbf{0.88}
    & \textbf{0.67}
    & \underline{0.89}
    & \textbf{0.35}
    & \textbf{0.80}
    & \textbf{0.16} \\
    
    BackFlip-2-joint
    & \underline{1.04}
    & \underline{0.82}
    & \textbf{0.90}
    & \textbf{0.35}
    & \underline{0.78}
    & \underline{0.17} \\ 

    \bottomrule
    \end{tabular}
    }%
    \label{tab:app_joint_trained_on_atlas}
\end{table}

\subsection{Ablation experiments}
\label{sec:app_denovo_metrics}

\begin{figure*}[h!]
    \centering
    \includegraphics[width=1.0\textwidth]{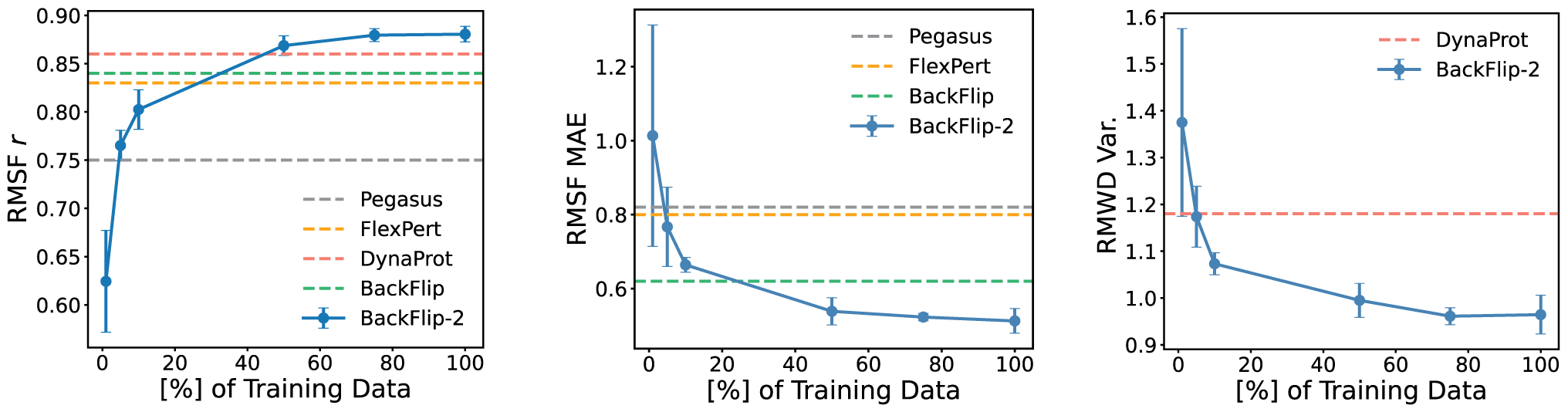}
    \caption{Learning curves on the ATLAS test set. Our model outperforms flexibility prediction baselines using only a fraction of the training data. We report RMSF correlation and MAE, and RMWD of per-residue covariances as a function of the training dataset size. For our model, each point reports the mean and standard deviation over 5 training runs on the ATLAS dataset.}
    \label{fig:learning_curve}
    
\end{figure*}

\begin{table*}[t]
    \centering
    \caption{Ablation of BackFlip-2. We focus on the equivariance of the per-residue covariance output and the Cholesky factorization from DynaProt~\cite{dynaprot} compared to the proposed simpler parametrization (Eq.~\ref{eq:local_cov}). All metrics are reported as medians over all proteins, analogous to Tab.~\ref{tab:covar_performance}. Best value is highlighted in bold, second-best is underlined.}
    \resizebox{\textwidth}{!}{%
    \begin{tabular}{lcc cccccc}
    \toprule
    \multirow{2}{*}{Model}
    & Equiv. 
    & Per-res
    & RMWD
    & Sym.\ KL
    & RMSF $r$
    & RMSF
    & DCCM $r$
    & DCCM \\
    & 
    & Cholesky
    & ($\downarrow$)
    & ($\downarrow$)
    & ($\uparrow$)
    & MAE ($\downarrow$)
    & ($\uparrow$)
    & MAE ($\downarrow$) \\
    \midrule
    Ours
    & \checkmark
    & $\times$
    & \textbf{0.88}
    & \textbf{0.67}
    & \underline{0.89}
    & \underline{0.35}
    & \textbf{0.80}
    & \textbf{0.16} \\
    \midrule
    a
    & $\times$
    & $\times$
    & 1.05
    & 1.06
    & 0.88
    & 0.38
    & 0.71
    & 0.20 \\
    b
    & \checkmark
    & \checkmark
    & \underline{0.93}
    & \underline{0.69}
    & \textbf{0.90}
    & \underline{0.35}
    & \underline{0.78}
    & \underline{0.17} \\
    c
    & $\times$
    & \checkmark
    & 1.05
    & 1.08
    & 0.88
    & \textbf{0.33}
    & 0.74
    & 0.19 \\
    \bottomrule
    \end{tabular}
    }
    \label{tab:app_model_ablation}
\end{table*}

\begin{table}[t]
    \centering
    \caption{Performance of BackFlip-2 when either distorted equilibrium or
    AlphaFold2-predicted structures are provided as input. We add Gaussian noise
    with the indicated standard deviation to the backbone atoms of equilibrium
    structures, and rerun inference with the model for the ATLAS test set. Metrics
    are reported as in Tab.~\ref{tab:covar_performance}. \textbf{Bold} marks the
    best and \underline{underline} the second-best value per column.}
    \vspace{0.1cm}
    \resizebox{\textwidth}{!}{%
    \begin{tabular}{lcccccc}
    \toprule
    Method 
    & RMWD Var. ($\downarrow$) 
    & KL Var. ($\downarrow$)
    & RMSF $r$ ($\uparrow$)
    & RMSF MAE ($\downarrow$)
    & DCCM $r$ ($\uparrow$)
    & DCCM MAE ($\downarrow$) \\
    \midrule
    BackFlip-2
    & \textbf{0.88}
    & \textbf{0.67}
    & \textbf{0.89}
    & \textbf{0.35}
    & \textbf{0.80}
    & \textbf{0.16} \\
    \midrule
    BackFlip-2 $+0.2$\,\AA{}
    & \underline{0.96}
    & \underline{0.71}
    & \underline{0.88}
    & \textbf{0.35}
    & \underline{0.79}
    & \textbf{0.16} \\
    BackFlip-2 $+0.5$\,\AA{}
    & 0.99
    & 0.82
    & \textbf{0.89}
    & 0.40
    & \textbf{0.80}
    & \underline{0.17} \\
    BackFlip-2 $+$ AF2
    & 0.99
    & 0.77
    & \textbf{0.89}
    & \underline{0.36}
    & 0.78
    & \underline{0.17} \\
    \bottomrule
    \end{tabular}
    }%
    \label{tab:app_noised_inputs}
\end{table}

\newpage
\subsection{Limitations to learn the dynamical descriptors derived from long-timescale MD simulations}
\label{sec:app_limitations}

There are two main obstacles for extending our method to long- (millisecond to second) timescales. The first challenge is of practical nature. MD is expensive, hence existing millisecond-MD datasets are limited and cover only a few small proteins \cite{lindorff2011fast}. Approaches for modeling long-timescale dynamics hence rely on alternative tasks and are not trained to reproduce MD-derived properties. For example, BioEmu \cite{bioemu} is trained on structural ensembles derived from experiment and related proteins from the AlphaFold database. In principle, our method could also be trained on such a dataset. However, the second obstacle is more fundamental: the dynamical descriptors that our method learns, namely the per-residue covariance and pairwise couplings, rely on the assumption that the ground truth ensemble can be approximated as a Gaussian distribution (see Sec. \ref{sec:targets_description}, and Eq. \ref{eq:full_cov}). This Gaussianity assumption, in most cases, breaks down for large conformational changes and alternative folding states, which typically occur on long timescales. Hence the observables that the model would learn could lose their meaning. Even if they were predicted correctly (for example via coarse-grained representations), they would not be suitable to describe the biophysics of the system. We hence see ensemble generation models like BioEmu as more suitable for long timescales, and our method (of directly predicting the observables that are often of interest) as more suitable for shorter, nanoseconds-microsecond timescales.

\subsection{Convergence of MD reference observables}
\label{sec:app_convergence}
\begin{figure*}[h]
    \centering
    \includegraphics[width=1.0\textwidth]{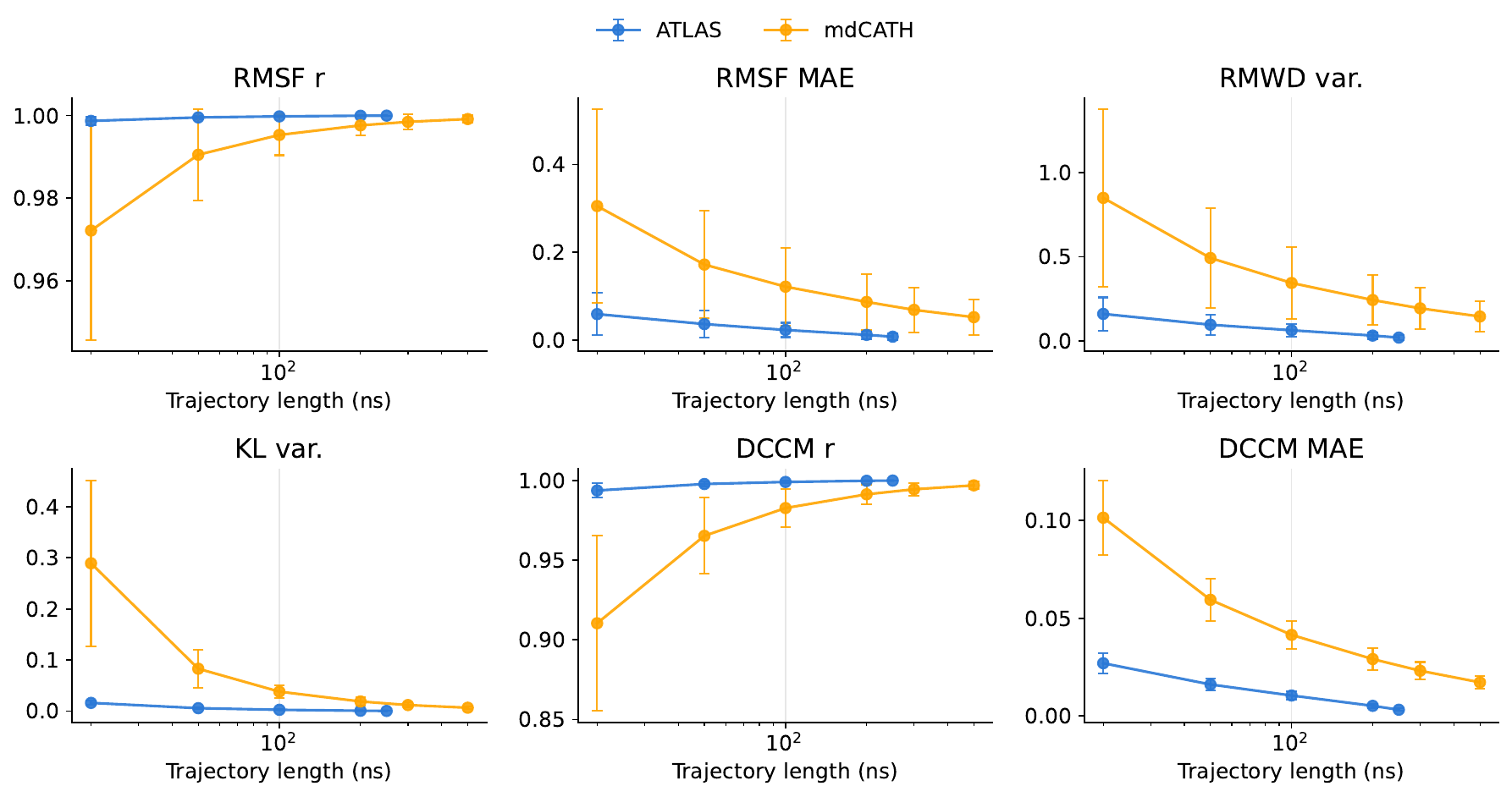}
    \caption{Convergence of the evaluation metrics on ATLAS and mdCATH. Each point
    compares observables computed from a truncated trajectory at the indicated length
    against those derived from the full trajectory (300\,ns for ATLAS, 500\,ns for mdCATH).}
    \label{fig:app_convergence_plot}
\end{figure*}

We evaluated the convergence of the metrics reported in Tab. \ref{tab:covar_performance} on both ATLAS and mdCATH. For each ATLAS test protein, we concatenated the three 100 ns trajectories and computed per-residue covariance matrices, RMSF, and DCCMs from the full 300 ns trajectory. We then randomly sampled five time windows for each length 20, 100, 200, and 250 ns to estimate the deviation between such truncated trajectories and the full 300ns trajectory. At each trajectory length, we report the full set of metrics relative to the complete 300 ns trajectory. We conducted the same computation with the mdCATH dataset using trajectory lengths of 20, 100, 200, 300, and 500 ns. We report the metrics in Fig. \ref{fig:app_convergence_plot}. 

For ATLAS, we find that the dynamical descriptors are effectively converged before reaching 300 ns. At 250 ns, the RMSF- and DCCM-related metrics are fully converged, and the RMWD variance error relative to the full trajectory is approximately 0.02. RMSF and DCCM, global properties of the system, converge faster than RMWD and the limit values are similar across the two datasets. Importantly, the ATLAS RMWD deviation decreases consistently towards zero with increasing trajectory length, indicating that it is convergent as well. Thus, the finite sampling noise is substantially smaller than the performance differences reported in Tab. \ref{tab:covar_performance}. In case of mdCATH, RMWD converges towards a systematically higher value compared to ATLAS, possibly suggesting that the discrepancy between datasets cannot be explained solely by insufficient sampling and most likely, indeed, reflects differences in simulation protocol, including the higher temperature (320K) and integration timestep (4 fs) used for mdCATH.


\end{document}